\documentclass[aps,prl,twocolumn,showpacs,superscriptaddress,nofootinbib]{revtex4}

\usepackage{amsmath}   
\usepackage{graphicx}  
\usepackage{xspace}
\usepackage{units}
\usepackage{lscape}

\usepackage{hyperref}

\newcommand{\minerva}{MINERvA\xspace}

\newcommand{\numubar}{\ensuremath{\bar{\nu}_{\mu}}\xspace}

\newcommand{\Eavail}{\ensuremath{E_\mathrm{avail}}\xspace}
\newcommand{\ud}{\ensuremath{\mathrm{d}}}
\newcommand{\dsdEdq}{\ensuremath{\ud^2\sigma/\ud \Eavail \ud q_{3}}\xspace}

\newcommand{\sizecheck}{0} 

\newif\ifpdf
\ifx\pdfoutput\undefined
   \pdffalse
\else
   \pdfoutput=1
   \pdftrue
\fi
\ifpdf
   \usepackage{graphicx}
   \usepackage{epstopdf}
\else
   \usepackage{graphicx}
\fi

\begin{document}

\title{Antineutrino Charged-Current Reactions on Hydrocarbon with Low Momentum 
  Transfer}





\newcommand{\deceased}{Deceased}


\newcommand{\Rutgers}{Rutgers, The State University of New Jersey, Piscataway, New Jersey 08854, USA}
\newcommand{\Hampton}{Hampton University, Dept. of Physics, Hampton, Virginia 23668, USA}
\newcommand{\Dortmund}{Institute of Physics, Dortmund University, 44221, Germany }
\newcommand{\Otterbein}{Department of Physics, Otterbein University, 1 South Grove Street, Westerville, Ohio 43081, USA}
\newcommand{\JMU}{James Madison University, Harrisonburg, Virginia 22807, USA}
\newcommand{\Florida}{University of Florida, Department of Physics, Gainesville, Florida 32611, USA}
\newcommand{\UCIrvine}{Department of Physics and Astronomy, University of California, Irvine, Irvine, California 92697-4575, USA}
\newcommand{\CBPF}{Centro Brasileiro de Pesquisas F\'{i}sicas, Rua Dr. Xavier Sigaud 150, Urca, Rio de Janeiro, Rio de Janeiro, 22290-180, Brazil}
\newcommand{\PUCP}{Secci\'{o}n F\'{i}sica, Departamento de Ciencias, Pontificia Universidad Cat\'{o}lica del Per\'{u}, Apartado 1761, Lima, Per\'{u}}
\newcommand{\INRM}{Institute for Nuclear Research of the Russian Academy of Sciences, 117312 Moscow, Russia}
\newcommand{\Jlab}{Jefferson Lab, 12000 Jefferson Avenue, Newport News, VA 23606, USA}
\newcommand{\Pittsburgh}{Department of Physics and Astronomy, University of Pittsburgh, Pittsburgh, Pennsylvania 15260, USA}
\newcommand{\Guanajuato}{Campus Le\'{o}n y Campus Guanajuato, Universidad de Guanajuato, Lascurain de Retana No. 5, Colonia Centro, Guanajuato 36000,  M\'{e}xico.}
\newcommand{\Athens}{Department of Physics, University of Athens, GR-15771 Athens, Greece}
\newcommand{\Tufts}{Physics Department, Tufts University, Medford, Massachusetts 02155, USA}
\newcommand{\WM}{Department of Physics, College of William \& Mary, Williamsburg, Virginia 23187, USA}
\newcommand{\FNAL}{Fermi National Accelerator Laboratory, Batavia, Illinois 60510, USA}
\newcommand{\Purdue}{Department of Chemistry and Physics, Purdue University Calumet, Hammond, Indiana 46323, USA}
\newcommand{\MCLA}{Massachusetts College of Liberal Arts, 375 Church
  Street, North Adams, Massachusetts 01247, USA}
\newcommand{\UMD}{Department of Physics and Astronomy, University of Minnesota -- Duluth, Duluth, Minnesota 55812, USA}
\newcommand{\Northwestern}{Northwestern University, Evanston, Illinois  60208, USA}
\newcommand{\UNI}{Universidad Nacional de Ingenier\'{i}a, Apartado 31139, Lima, Per\'{u}}
\newcommand{\Rochester}{University of Rochester, Rochester, New York 14627 USA}
\newcommand{\Austin}{Department of Physics, University of Texas, 1 University Station, Austin, Texas 78712, USA}
\newcommand{\USM}{Departamento de F\'{i}sica, Universidad T\'{e}cnica Federico Santa Mar\'{i}a, Avenida Espa\~{n}a 1680 Casilla 110-V, Valpara\'{i}so, Chile}
\newcommand{\Geneva}{University of Geneva, 1211 Geneva 4, Switzerland}
\newcommand{\Chicago}{Enrico Fermi Institute, University of Chicago, Chicago, IL 60637 USA}
\newcommand{\hired}{}
\newcommand{\OregonState}{Department of Physics, Oregon State University, Corvallis, Oregon 97331, USA}
\newcommand{\oxford}{Oxford University, Department of Physics, Oxford, United Kingdom}
\newcommand{\umiss}{University of Mississippi, Oxford, Mississippi 38677, USA}
\newcommand{\upenn}{University of Pennsylvania, Philadelphia, Pennsylvania 19104, USA}
\newcommand{\AMU}{AMU Campus, Aligarh, Uttar Pradesh 202001, India}
\newcommand{\wroclaw}{University of Wroclaw, plac Uniwersytecki 1, 50-137 Wrocław, Poland}
\newcommand{\Mohali}{Knowledge city, Sector 81, SAS Nagar, Manauli PO 140306}
\newcommand{\chrismarshallThanks}{now at Lawrence Berkeley National Laboratory, Berkeley, CA 94720, USA}
\newcommand{\cpatrickThanks}{Now at University College London, London WC1E 6BT, UK}
\newcommand{\jwolcottThanks}{now at Tufts University, Medford, MA 02155, USA}
\newcommand{\melkinsThanks}{Now at Iowa State University, Ames, IA 50011, USA }
\newcommand{\joelmousseauThanks}{now at University of Michigan, Ann Arbor, MI 48109, USA}


\author{R.~Gran}                          \affiliation{\UMD}
\author{M.~Betancourt}                    \affiliation{\FNAL}
\author{M.~Elkins}\thanks{\melkinsThanks}                     \affiliation{\UMD}
\author{P.A.~Rodrigues}          \affiliation{\oxford}  \affiliation{\umiss}     \affiliation{\Rochester}


\author{F.~Akbar}                         \affiliation{\AMU}
\author{L.~Aliaga}                        \affiliation{\WM}  \affiliation{\PUCP}
\author{D.A.~Andrade}                     \affiliation{\Guanajuato}
\author{A.~Bashyal}  \affiliation{\OregonState}
\author{L.~Bellantoni}                    \affiliation{\FNAL}
\author{A.~Bercellie}                     \affiliation{\Rochester}
\author{A.~Bodek}                         \affiliation{\Rochester}
\author{A.~Bravar}                        \affiliation{\Geneva}
\author{H.~Budd}                          \affiliation{\Rochester}
\author{G.F.R.~Caceres~Vera}            \affiliation{\CBPF}
\author{T.~Cai}                           \affiliation{\Rochester}
\author{M.F.~Carneiro}                    \affiliation{\OregonState}
\author{D.~Coplowe}                       \affiliation{\oxford}
\author{H.~da~Motta}                      \affiliation{\CBPF}
\author{S.A.~Dytman}                      \affiliation{\Pittsburgh}
\author{G.A.~D\'{i}az~}                   \affiliation{\Rochester}  \affiliation{\PUCP}
\author{J.~Felix}                         \affiliation{\Guanajuato}
\author{L.~Fields}                        \affiliation{\FNAL}  \affiliation{\Northwestern}
\author{R.~Fine}                          \affiliation{\Rochester}
\author{H.~Gallagher}                     \affiliation{\Tufts}
\author{A.~Ghosh}                         \affiliation{\USM}  \affiliation{\CBPF}
\author{H.~Haider}                        \affiliation{\AMU}
\author{J.Y.~Han}                         \affiliation{\Pittsburgh}
\author{D.A.~Harris}                      \affiliation{\FNAL}
\author{S.~Henry}                         \affiliation{\Rochester}
\author{D.~Jena}                           \affiliation{\FNAL}
\author{J.~Kleykamp}                      \affiliation{\Rochester}
\author{M.~Kordosky}                      \affiliation{\WM}
\author{T.~Le}                            \affiliation{\Tufts}  \affiliation{\Rutgers}
\author{J.R.~Leistico}  \affiliation{\UMD}
\author{A.~Lovlein}  \affiliation{\UMD}  
\author{X.-G.~Lu}                         \affiliation{\oxford}
\author{E.~Maher}                         \affiliation{\MCLA}
\author{S.~Manly}                         \affiliation{\Rochester}
\author{W.A.~Mann}                        \affiliation{\Tufts}
\author{C.M.~Marshall}  \thanks{\chrismarshallThanks}  \affiliation{\Rochester}
\author{K.S.~McFarland}                   \affiliation{\Rochester}  \affiliation{\FNAL}
\author{A.M.~McGowan}                     \affiliation{\Rochester}
\author{B.~Messerly}                      \affiliation{\Pittsburgh}
\author{J.~Miller}                        \affiliation{\USM}
\author{A.~Mislivec}                      \affiliation{\Rochester}
\author{J.G.~Morf\'{i}n}                  \affiliation{\FNAL}
\author{J.~Mousseau}\thanks{\joelmousseauThanks}  \affiliation{\Florida}
\author{D.~Naples}                        \affiliation{\Pittsburgh}
\author{J.K.~Nelson}                      \affiliation{\WM}
\author{C.~Nguyen}  \affiliation{\Florida}
\author{A.~Norrick}                       \affiliation{\WM}
\author{Nuruzzaman}                       \affiliation{\Rutgers}  \affiliation{\USM}
\author{A.~Olivier}                       \affiliation{\Rochester}
\author{V.~Paolone}                       \affiliation{\Pittsburgh}
\author{C.E.~Patrick}\thanks{\cpatrickThanks}  \affiliation{\Northwestern}
\author{G.N.~Perdue}                      \affiliation{\FNAL}  \affiliation{\Rochester}
\author{M.A.~Ram\'{i}rez}                 \affiliation{\Guanajuato}
\author{R.D.~Ransome}                     \affiliation{\Rutgers}
\author{H.~Ray}                           \affiliation{\Florida}
\author{L.~Ren}                           \affiliation{\Pittsburgh}
\author{D.~Rimal}                         \affiliation{\Florida}
\author{D.~Ruterbories}                   \affiliation{\Rochester}
\author{H.~Schellman}                     \affiliation{\OregonState}  \affiliation{\Northwestern}
\author{C.J.~Solano~Salinas}              \affiliation{\UNI}
\author{H.~Su}                            \affiliation{\Pittsburgh}
\author{M.~Sultana}                       \affiliation{\Rochester}
\author{S.~S\'{a}nchez~Falero}            \affiliation{\PUCP}
\author{E.~Valencia}                      \affiliation{\WM}  \affiliation{\Guanajuato}
\author{J.~Wolcott}\thanks{\jwolcottThanks}  \affiliation{\Rochester}
\author{M.~Wospakrik}                      \affiliation{\Florida}
\author{B.~Yaeggy}                        \affiliation{\USM}

%
\collaboration{\minerva  Collaboration}\ \noaffiliation


\begin{abstract}
We report on multinucleon effects in low momentum transfer ($< 0.8$ GeV/c) 
antineutrino interactions on plastic (CH) scintillator.   These data are
from the 2010-2011 antineutrino phase of the MINERvA experiment at
Fermilab.   The hadronic energy spectrum of this
inclusive sample is well described when 
a screening effect at low energy transfer and a two-nucleon knockout
process are added to a relativistic Fermi gas model
of quasielastic, $\Delta$ resonance, and higher resonance processes.
In this analysis, model elements introduced to describe previously
published neutrino results have quantitatively similar benefits for
this antineutrino sample.
We present the results as a double-differential cross section to
accelerate the investigation of alternate models for antineutrino scattering off nuclei.
\end{abstract}
\ifnum\sizecheck=0
\maketitle
\fi


Current and future accelerator-based neutrino oscillation
experiments analyze flavor oscillations based on distortions of
reconstructed antineutrino energy spectra.
These measurements require models for both the
lepton energy and angle, and for the hadronic system.
Experiments using calorimetric reconstruction \cite{Adamson:2016xxw, Adamson:2016tbq}
are especially sensitive to the presence of 
neutrons in the final state.
To probe for charge-parity (CP) violation in the lepton sector~\cite{ Acciari:2015zzz,
Adams:2013qkq,Acciarri:2015uup}, 
models of antineutrino processes require
similar accuracy to the corresponding neutrino processes.
Otherwise, model uncertainties limit the sensitivity to,
or possibly mimic, a CP-violating effect.

We present the first antineutrino analysis of inclusive charged-current
reactions to isolate multinucleon effects in the
quasielastic (CCQE) and $\Delta$ resonance
kinematic regions.
We reconstruct the hadronic system using calorimetry and we obtain an
estimate of the three-momentum transfer for each event.
The data are subdivided into six
subranges of momentum transfer up to 0.8 GeV/c, and within each range we present the
observed hadronic energy in the detector.
To describe these data, a component of the event rate could be
attributed to
many-body effects like a two-particle,
two-hole (2p2h) process~\cite{Rodrigues:2015hik, Martini:2009uj, Nieves:2011pp,
  Gonzalez-Jimenez:2014eqa, Megias:2016fjk, VanCuyck:2017wfn,
  RuizSimo:2017hlc, Lightbody:1988, Sobczyk:2012ms, Schwehr:2016pvn, 
  Gallmeister:2016dnq}.
Also, suppression of CCQE interactions is preferred, 
such as provided by a random phase approximation (RPA) calculation
\cite{Nieves:2004wx,Martini:2009uj,Pandey:2014tza,Nieves:2017lij} applied to a Fermi
gas model \cite{Smith:1972xh}.

The data were taken between November 2010 and February 2011
with the NuMI beam~\cite{Adamson:2015dkw} operating in antineutrino mode.
The primary beam of 120-GeV protons interacts in a
graphite target, producing mesons.   A pair of magnetic horns focuses
negatively-charged mesons toward a decay pipe where their decay leads
to an antineutrino spectrum in the
MINERvA detector peaking near 3.0 GeV.   
We use a
G{\small EANT}4-based \cite{Agostinelli2003250,1610988} 
prediction for the flux with central values and
uncertainties adjusted~\cite{Aliaga:2016oaz} using thin-target
hadron production data~\cite{Alt:2006fr, Denisov:1973zv,
  Carroll:1978hc,Allaby:1969de} and an in situ neutrino-electron
scattering constraint~\cite{Park:2015eqa}.

A sample of charged-current \numubar interactions are selected
from MINERvA's 5.3-ton fiducial volume by requiring that a muon track leaves the MINERvA
detector and has its positive charge and momentum identified 
in the MINOS magnetized iron spectrometer~\cite{Michael:2008bc} located 2 m downstream.
The fiducial volume is both an active tracker and a
calorimeter, built from planes of scintillator (CH) strips with a
triangular shaped 
3.3-cm base and 1.7-cm height.  Alternating
and nesting the triangles gives light-sharing information
that improves tracking resolution.   Each hexagonal
plane contains 127 strips up to 245-cm in length.   The
planes are installed with strips oriented vertically or rotated  $\pm 60^{\circ}$,
ensuring the precise reconstruction of the interaction point and muon
track angle, even when hadronic activity partially obscures the muon.  
The target mass consists of
8.2\%, 88.5\%, and 2.5\% hydrogen, carbon, and
oxygen, respectively, plus small amounts of heavier nuclei. 

Particles leaving
the active tracking region pass into the
electromagnetic calorimeter (ECAL) where thin sheets of lead are epoxied
to each scintillator plane.  Farther downstream are layers of hadronic
calorimetry using alternating planes of scintillator and passive steel.   
The calorimetric
and tracking capabilities of MINERvA are constrained relative to
G{\small EANT}4 v.9.4.p2 (with the Bertini Cascade option) using in-situ
\cite{Aliaga:2013uqz} and hadron test beam measurements
\cite{Aliaga:2015aqe}.
With no test beam measurements, the neutron response and its
uncertainties come
after adjusting the cross section to match the data
from \cite{Abfalterer:2001gw} as used by later versions of G{\small EANT}4.

The kinematics of each event are reconstructed using the measured muon
energy and angle, and measured energy deposits attributed to
hadrons.   The technique is nearly identical to~\cite{Rodrigues:2015hik}.
A full simulation of the reconstructed sample with calibrated detector response
is made using the G{\small EANT}4 simulation and {\small GENIE} version
2.8.4 neutrino event generator \cite{Andreopoulos201087}.  This
simulation is used to obtain a correction \cite{Devan:2015uak}, as a function of the 
calorimetrically measured hadronic
energy, to estimate the energy transfer $q_0$.  This correction is applied
identically to reconstructed simulation and data.   In both
cases, the calorimetric neutrino
energy estimate is $E_\nu = E_\mu + q_0$, where $E_\mu$ includes the
muon rest mass $M_\mu$.  The square of the four-momentum transfer is
$-q^2 = Q^2 =
2E_\nu(E_\mu - p_\mu\cos\theta_\mu) - M^2_\mu$,  and the
three-momentum 
transfer is simply $q_3 = \sqrt{Q^2 + q_0^2}$.   In this
Letter, the kinematics of the analysis sample are limited to $q_3
< 0.8$ GeV/c.   There are no other requirements on
reconstructed hadronic topologies for this inclusive sample.

The measured energy deposits are used
to form another calorimetric estimator, the available
  energy \Eavail \cite{Rodrigues:2015hik}.  This is
energy due to particles that deposit most or all of their energy
in the detector:  proton kinetic energy, charged pion kinetic energy,
electrons, positrons, and photons, including those from neutral pion
and eta decays.  These momentum transfers are too low for the production
of heavier mesons and baryons.


When \Eavail is formed from a model, it does not include neutrons
that leave a small fraction of their energy in the detector or the energy
used to unbind nucleons.   In the neutrino case~\cite{Rodrigues:2015hik},
where outgoing protons far outnumber neutrons, this is a good
approximation.  In the simulation of this antineutrino subsample,
70\% of interactions have more than half the energy transfer going to neutrons,
including 40\% which have neutron-only final states. 
Up to 60\% of neutrons at these energies leave reconstructed energy deposits in the
detector, so neutrons can contribute
significantly 
to the hadronic energy deposits.  Despite this, reconstructed and
model
distributions of \Eavail vs. $q_3$ retain the ability to separate CCQE
and $\Delta$ resonance kinematics and the region between them.
Because the analysis is limited to interactions with
little energy in the recoil system, only the energy deposits in the
tracker and downstream ECAL regions are considered.  
The backgrounds from unrelated beam activity are higher and calorimetric resolution is worse
for energy deposits in the other 
regions, degrading the sensitivity to multinucleon effects.

\begin{figure}[ht!]
\begin{center}
\includegraphics[width=8.7cm]{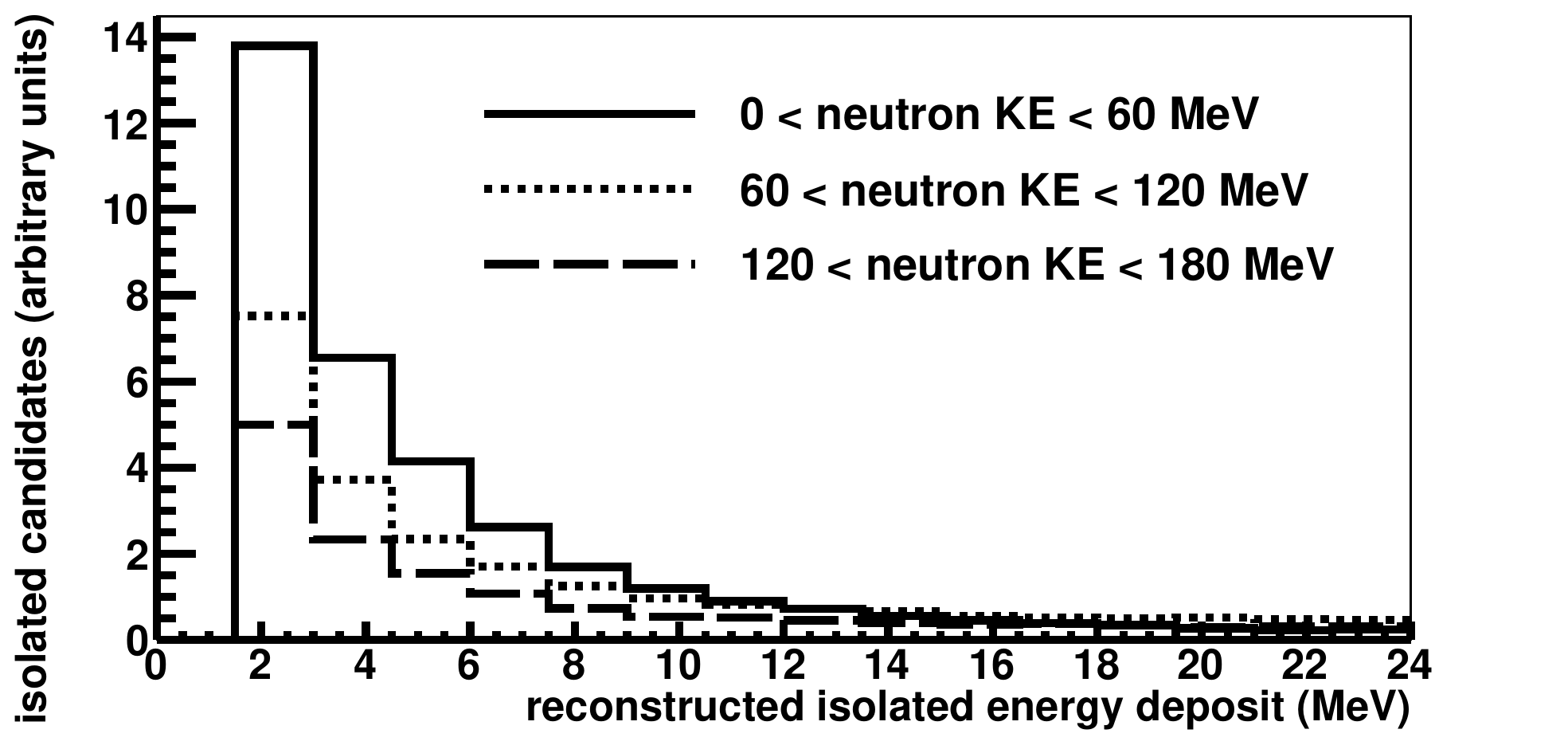}
\caption{Predicted isolated energy deposits from neutrons in three ranges of neutron
  kinetic energy (KE).    These are selected from the tuned MC sample
  (described later)
  with $q_3 < 0.8$ GeV/c.  The three curves also illustrate the relative
  abundance of lower kinetic energies in the selected MC sample.
  Neutrons in this energy range typically leave small
  isolated energy deposits, uncorrelated with the neutron
  kinetic energy.
\label{fig:neutronedep}}
\end{center}
\end{figure}

While \Eavail is defined assuming neutrons have a negligible
calorimetric response, the actual situation is more complex.  
Interactions that have only neutrons in the final state are most
likely to have reconstructed hadronic energy between 0 and 10 MeV.
Figure~\ref{fig:neutronedep} shows the reconstructed
energy deposits from the {\small GENIE}-produced neutrons exiting the nucleus
and simulated by G{\small EANT}4 with the detector model.
The most common outcomes are small ($<$~10~MeV) energy deposits as the
neutrons scatter on hydrogen and carbon in the detector.  The
response is mostly uncorrelated with the neutron energy, and it is not
suited for a calorimetric quantity.  Larger secondary proton energy
deposits become more common as neutron kinetic energy increases.  For this analysis and its
kinematic range,
neutrons with tens to a few hundreds of MeV are effectively
treated as biasing reconstructed \Eavail to higher values than the
true quantity.

\begin{figure}[ht!]
\begin{center}
\includegraphics[width=8.0cm]{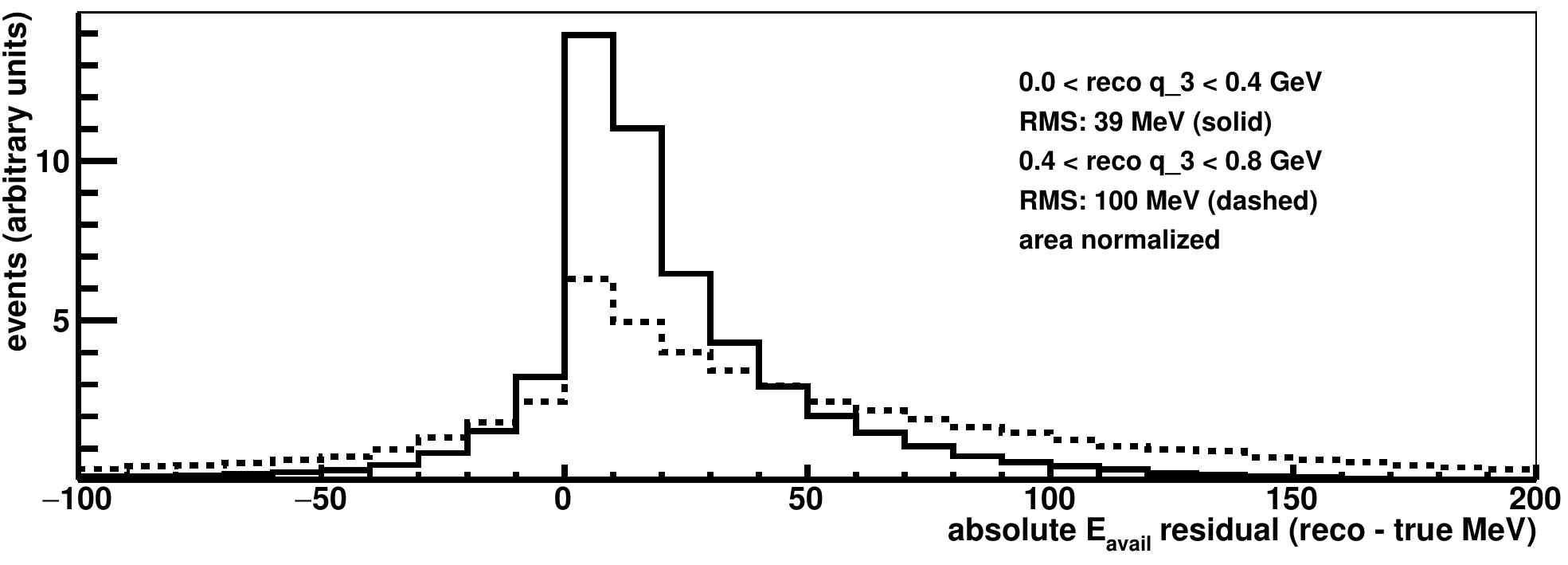}
\includegraphics[width=8.0cm]{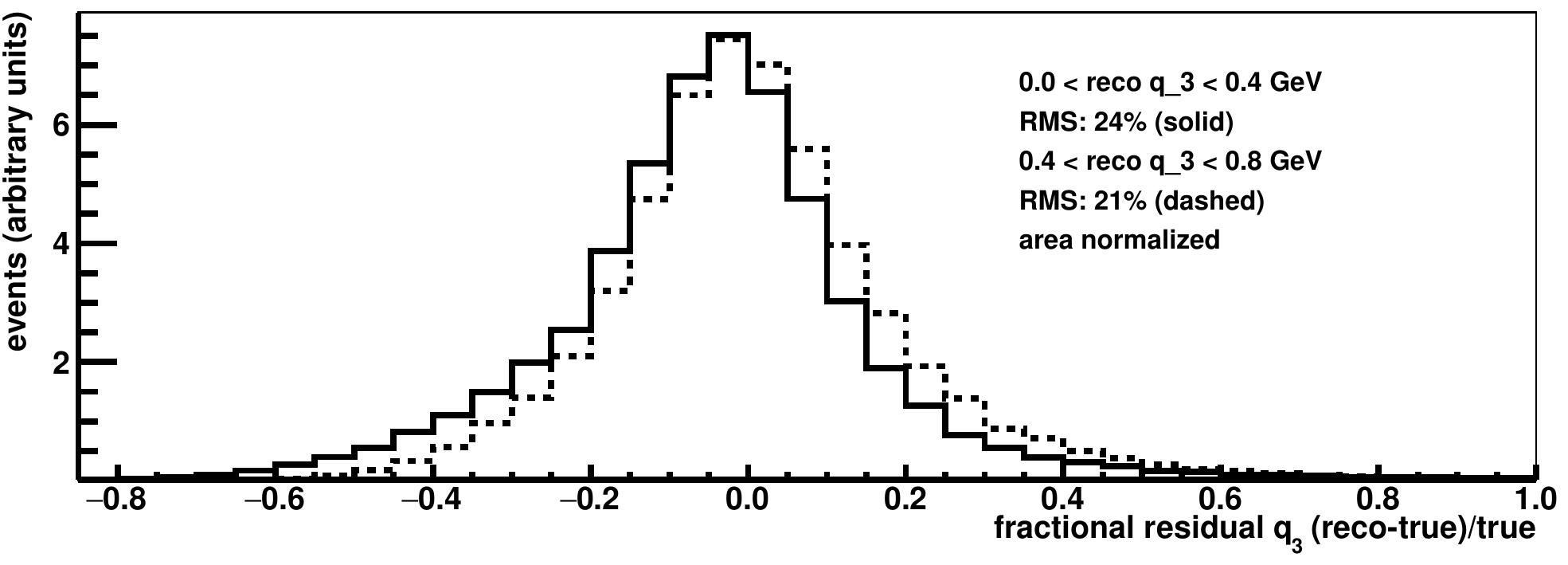}
\caption{Absolute resolution of \Eavail (top) and fractional
  resolution for three-momentum transfer (bottom), for the MC sample
  with reconstructed $q_3 < 0.8$ GeV/c.
  The dashed line is the upper half
  of the $q_3$ range; the solid line is the lower half.
\label{fig:resolution}}
\end{center}
\end{figure}

Another consequence of the neutron response in the MINERvA detector is that the
resolutions for some reconstructed quantities are different
than the neutrino case.  The two hadronic energy estimators for
the selected sample have
significantly worse resolutions.  The simulation indicates
a root-mean-square (rms) resolution of  58\% for $q_0$ compared to
51\% for the neutrino case in \cite{Rodrigues:2015hik}.   For \Eavail the rms is also 58\%
while the neutrino resolution is significantly improved
to 40\%.   When neutrons
may be the only final state particle, the absolute residual is a
better metric (shown in Fig.~\ref{fig:resolution}) than the fractional rms.
The neutrino
energy estimator is negligibly different; because these events had
such little hadronic energy to begin with, the muon energy dominates the
resolution.
Muon energy and angle drive $q_3$.  Its resolution is barely degraded from 22\% to 23\% rms 
and it varies little
across the range of $q_3$.

Because this is an analysis of an inclusive sample, event selection is
minimal.  We only create a boundary for unfolding the data into a double
differential cross section that can be reproduced by
external event generators, and we exclude regions of kinematic space
that do not have good acceptance. 
The muon momentum is required to be
above 1.5 GeV/c and an angle less than 20 degrees with respect to the beam
direction. 
We further limit the reconstructed antineutrino energy
to between 2 and 6 GeV, which spans the peak of this beam and
allows a direct comparison to the neutrino results
\cite{Rodrigues:2015hik}.
These selections are used for the reconstructed events, the unfolded
distribution, and the true distribution of MC simulations compared to the latter.

\begin{figure}[t]
\begin{center}
\includegraphics[width=8.4cm]{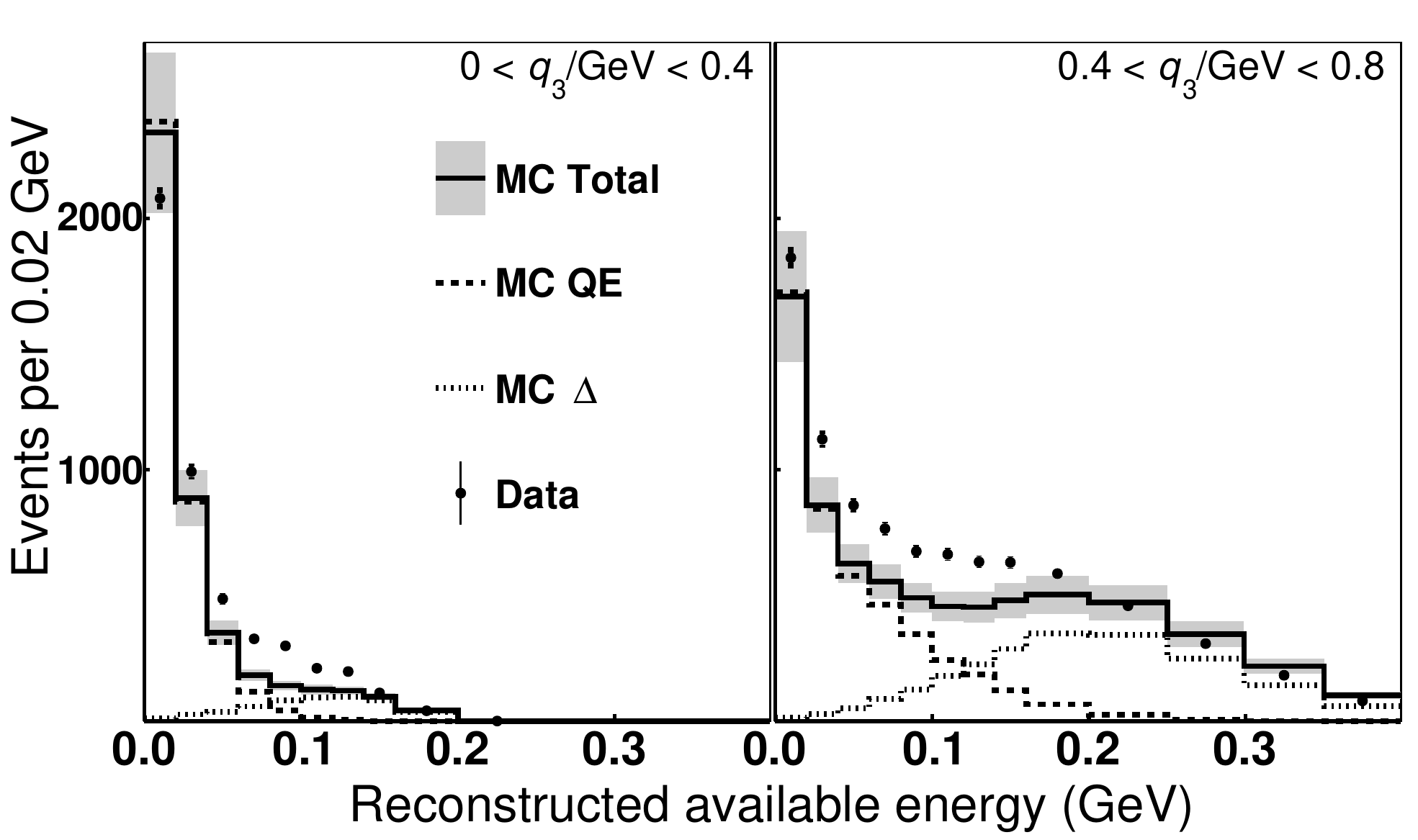}
\includegraphics[width=8.4cm]{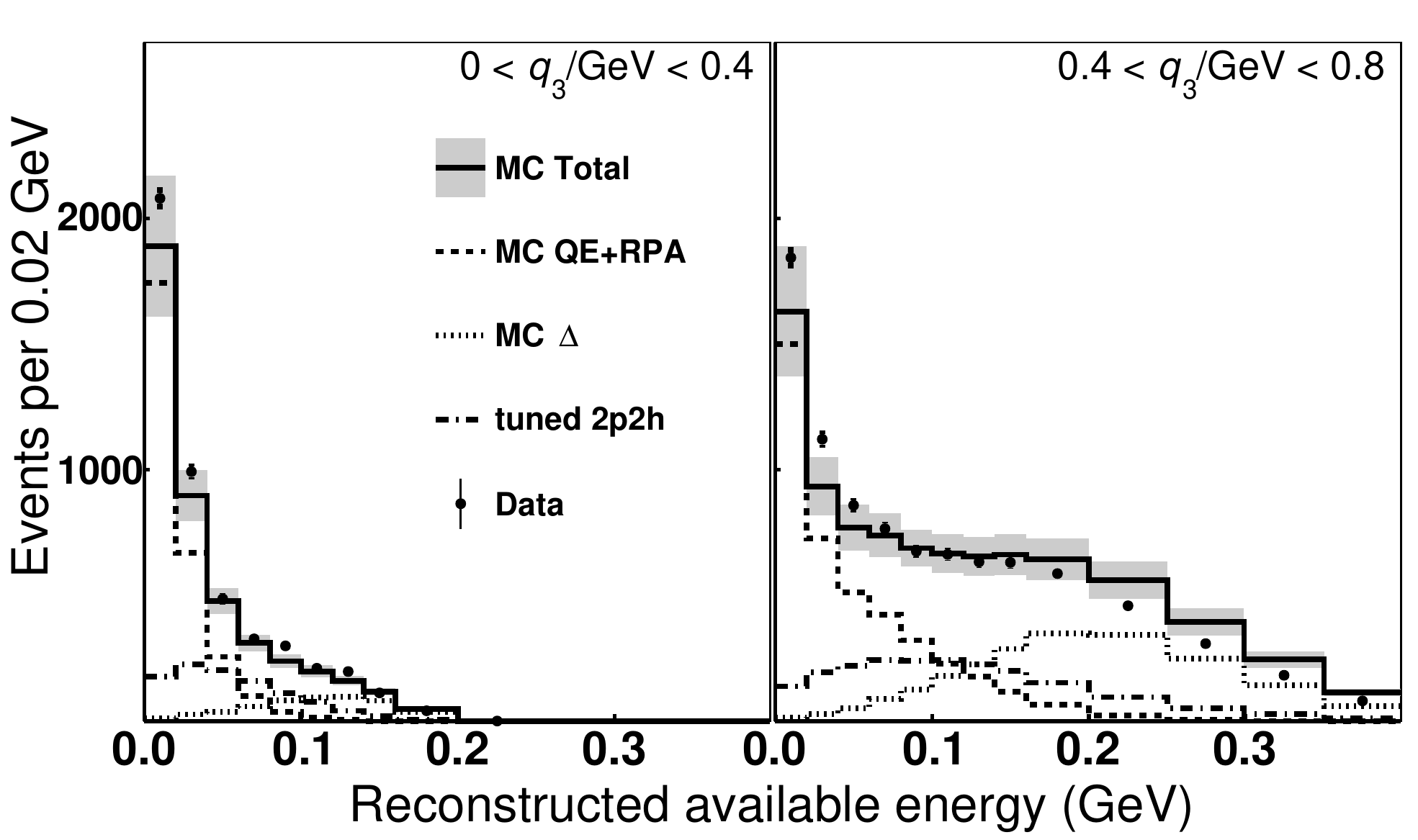}
\caption{Reconstructed \Eavail distributions compared to (top) the
  base Monte Carlo simulation (GENIE with minor modifications to pion
  production) for two ranges of reconstructed three
  momentum transfer.  In the improved simulation MnvGENIE-v1 (bottom),
the 
region between the predicted CCQE process (dashed) and the
$\Delta$(1232) resonance (dotted) is filled by
events generated from the Valencia 2p2h process
plus additional 2p2h events (dot-dashed).
\label{fig:reco2}}
\end{center}
\end{figure}

The selected inclusive sample is compared to the prediction of the
GENIE event generator combined with a G{\small EANT}4 simulation of the
outgoing particles from the reaction.  GENIE's simulation of the CCQE
process is from Llewellyn Smith \cite{LlewellynSmith:1971zm} with vector form factors
parameterized by \cite{Bradford:2006yz}, and the axial form factor is
taken to be a dipole with an
axial mass of 0.99 GeV.  For interactions on carbon and other nuclei,
GENIE uses a Fermi gas model \cite{Smith:1972xh}.   The $\Delta$ and
higher resonances use Rein and Sehgal \cite{Rein:1980wg}, with a nonresonant
component taken from the deeply inelastic scattering model~\cite{Bodek:2004pc}
as the resonances are phased out from invariant
mass 1.4 $<$ W $<$ 2.0 GeV. 
We add two minor (for this analysis $<2$\% of the total rate)
modifications to pion production.  The non-resonance, 
single-pion process is reduced to 43\% of the nominal following the comparison of GENIE to
bubble chamber experiment neutrino data \cite{Rodrigues:2016xjj,Wilkinson:2014yfa}.
Coherent pion events with pion kinetic energy $<$ 0.45 GeV are reduced
by half \cite{Higuera:2014azj,Mislivec:2017qfz,Berger:2008xs}.
This base combination of models, compared to reconstructed data, has
discrepancies in the region between
the CCQE and $\Delta$ process as large as a factor of two, as shown in
the top panels of Fig.~\ref{fig:reco2}.

We have modified the default GENIE version 2.8.4 to include advances
in modeling the important processes.  
The CCQE process is modified to include RPA screening based on 
the IFIC Valencia model
\cite{Nieves:2004wx,Gran:2013kda} implemented by weighting GENIE CCQE
events \cite{Gran:2017psn}. 
A CCQE-like two-particle, two-hole process ``2p2h'' from the
model by the same group \cite{Nieves:2011pp,Gran:2013kda}
is implemented in GENIE \cite{Schwehr:2016pvn}.

The IFIC Valencia 2p2h model increases the predicted event rates, but not enough.  This process
is increased further with an empirical
enhancement  \cite{minerva2p2h:2019}  based on MINERvA
inclusive neutrino
data \cite{Rodrigues:2015hik}.   The additional events are from weighting up
the generated 2p2h events according to a two-dimensional Gaussian in true $q_0,q_3$
whose six parameters
are fit to the neutrino data version of these distributions.  This
enhancement adds 50\% to the predicted 2p2h strength, but targets
the event rate in the kinematic region between the CCQE and $\Delta$
peaks where the rate doubles.
The collection of changes in this and the preceding paragraphs are
referred to as ``MnvGENIE-v1'' and are the central, tuned model
for many recent analyses ~\cite{Betancourt:2017uso, Altinok:2017xua, Patrick:2018gvi}.

The resulting description of the antineutrino data
is much improved, as illustrated in Fig.~\ref{fig:reco2} and summarized  in
Table~\ref{tab:chi2}
using a standard $\chi^2$ test on the reconstructed samples.
These models also improve the description of muon-only kinematic
distributions of an overlapping subset of the same
data set~\cite{Patrick:2018gvi} selected with no pions in the final state.

For this model comparison to reconstructed data, 
the largest systematic uncertainties include flux, 
hadron energy scale, and GENIE resonance interaction and final-state
rescattering model uncertainties.   The GENIE uncertainty on
the CCQE axial form factor is reduced to $\pm$9\% following the
analysis of \cite{Meyer:2016oeg}.  An 
uncertainty on the RPA CCQE suppression
\cite{Valverde:2006zn,Gran:2017psn} is added, most significantly from comparison to
muon capture data.   No single uncertainty dominates the model prediction
for the reconstructed distributions.

The antineutrino sample retains a discrepancy just beyond the error
band in the four second-lowest \Eavail bins within the range 0.3 $<$ $q_3$ $<$ 0.8 GeV/c.  
These bins are dominated by events with neutron-only
final states, including feed-down from higher energy transfer CCQE and 2p2h
reactions.   
Limited to the models available for this analysis, 
both the CCQE RPA and the tuned 2p2h component 
each have a 10\% to 30\% effect on these bins.
The comparison of the first two rows of Table \ref{tab:chi2} is subtle; the first
does not contain additional uncertainty from the RPA model.
Applying an estimate for the uncertainty to both rows also yields
worse $\chi^2$ for the lower $q_3$ range for
neutrino when RPA is added.
RPA reduces some
bins where the MC simulation is already under predicting the data.
However,  the RPA model produces a
better agreement in the lowest \Eavail for 0.0 $<$ $q_3$ $<$ 0.3 GeV/c,
which is also where the predicted
RPA effect is more significant than the predicted 2p2h effect. 
These data appear 
sensitive to details of the CCQE vs. 2p2h processes not yet exposed within
the available models, details such as those \cite{Pandey:2014tza,Martini:2016eec,Nieves:2017lij} that go beyond the Fermi gas.

This 2p2h tune comes with three other variations that treat the final state nucleon
content as uncertain.  Instead of enhancing all 2p2h
events, the first variation enhances only those generated for {\it pn} initial state nucleon
pairs,  which translates to {\it pp} final states for the neutrino case in the fit
and {\it nn} for the antineutrino case where we apply the tuned
parameters.   The next variation enhances reactions that are not on {\it pn}
initial state pairs, which lead to {\it pn} final states.  Finally,
the third variation enhances CCQE events at these kinematics. 
In addition to testing these variations against the reconstructed
data, they are used as an uncertainty applied
later when producing a double-differential cross section.

This sample also includes a significant component at and beyond the
$\Delta$ resonance peak, which remains poorly described by these model
variations.    The shortcoming of the model for these low $Q^2 =
q_3^2-q_0^2 \approx 0$ events shows up on the far right of the distributions in Fig.~\ref{fig:reco2}.
Similar mismodeling of the resonance-region rate has been previously reported in
measurements on mineral oil by MiniBooNE
\cite{AguilarArevalo:2010bm,AguilarArevalo:2010xt},
in MINERvA's pion final state samples
\cite{Altinok:2017xua,McGivern:2016bwh,Eberly:2014mra}, 
in the neutrino version of this analysis \cite{Rodrigues:2015hik},
and in a
resonance-rich neutrino+Fe sample from MINOS \cite{Adamson:2014pgc}.
The latter used a calorimetric sample as a sideband and tuned an {\it ad hoc}, low $Q^2$
suppression to the data in order to improve the
estimate of the resonance background in their CCQE analysis.
At $Q^2=0$ the rate is 40\% of nominal and becomes no
suppression by $Q^2=0.7$ (GeV/c)$^2$.
Applying the MINOS parameterization improves the description of 
these MINERvA data for some of those bins
at high $q_3$, but the
suppression goes too
far and produces a model deficit in the highest energy bins
of the low $q_3$ panel.  These bins in Fig.~\ref{fig:reco2} were
already well described, and the $\chi^2$ reflects that the agreement worsens.  
Either the single-parameter $Q^2$ weight 
or the tuning to neutrino+Fe data is not adequate to describe
the two dimensional kinematics of these
antineutrino+CH samples.


\begin{table}[ht!]
\begin{center}
\begin{tabular}{l|cc|cc}
sample  & \numubar & \numubar & $\nu_\mu$& $\nu_\mu$ \\
$q_3$ range & Lower & Upper & Lower & Upper \\ 
degrees of freedom & 19 & 37 & 24 & 41 \\ \hline
GENIE 2.8.4+pion \cite{Andreopoulos201087,Rodrigues:2016xjj,Wilkinson:2014yfa, Higuera:2014azj,Mislivec:2017qfz} &  239 & 167  & 437 & 281 \\
+QERPA \cite{Valverde:2006zn,Gran:2017psn}  & 261 & 140 & 265 & 253 \\
+2p2h \cite{Nieves:2011pp,Gran:2013kda}   & 105 & 108 & 149 & 294\\
+tune \cite{Rodrigues:2015hik, minerva2p2h:2019}  & 69 & 80 & 77 & 150 \\ \hline
tune only {\it pn} initial state & 65  & 86  & 76 & 160 \\
tune not {\it pn} initial state & 71 & 74  & 84 & 163 \\
tune CCQE reactions & 59 & 123 & 108 &  166 \\ \hline
+MINOS resonance tune \cite{Adamson:2014pgc} & 151 & 45 & 114 & 141 \\ \hline
\end{tabular}
\caption{Comparison of the models to reconstructed data showing the 
  evolution of the $\chi^2$ with each model change.  The
  reconstructed data and the base model are as in the top panels,
  and the``+tune'' model are as in the lower panels of Fig.~\ref{fig:reco2}.
  The calculation actually uses the resolution-driven six bins
  of $q_3$ \{0.0,0.2,0.3,0.4,0.5,0.6,0.8\} GeV/c for best sensitivity,
  and they are summed into the same two ranges shown in
  Fig.~\ref{fig:reco2}.
  The right-most columns are made using the neutrino data \cite{Rodrigues:2015hik}
  though the models being tested in this Letter have advanced since
  that earlier publication. 
\label{tab:chi2}}
\end{center}
\end{table}

\begin{figure}[th!]
\begin{center}
\includegraphics[width=8.7cm]{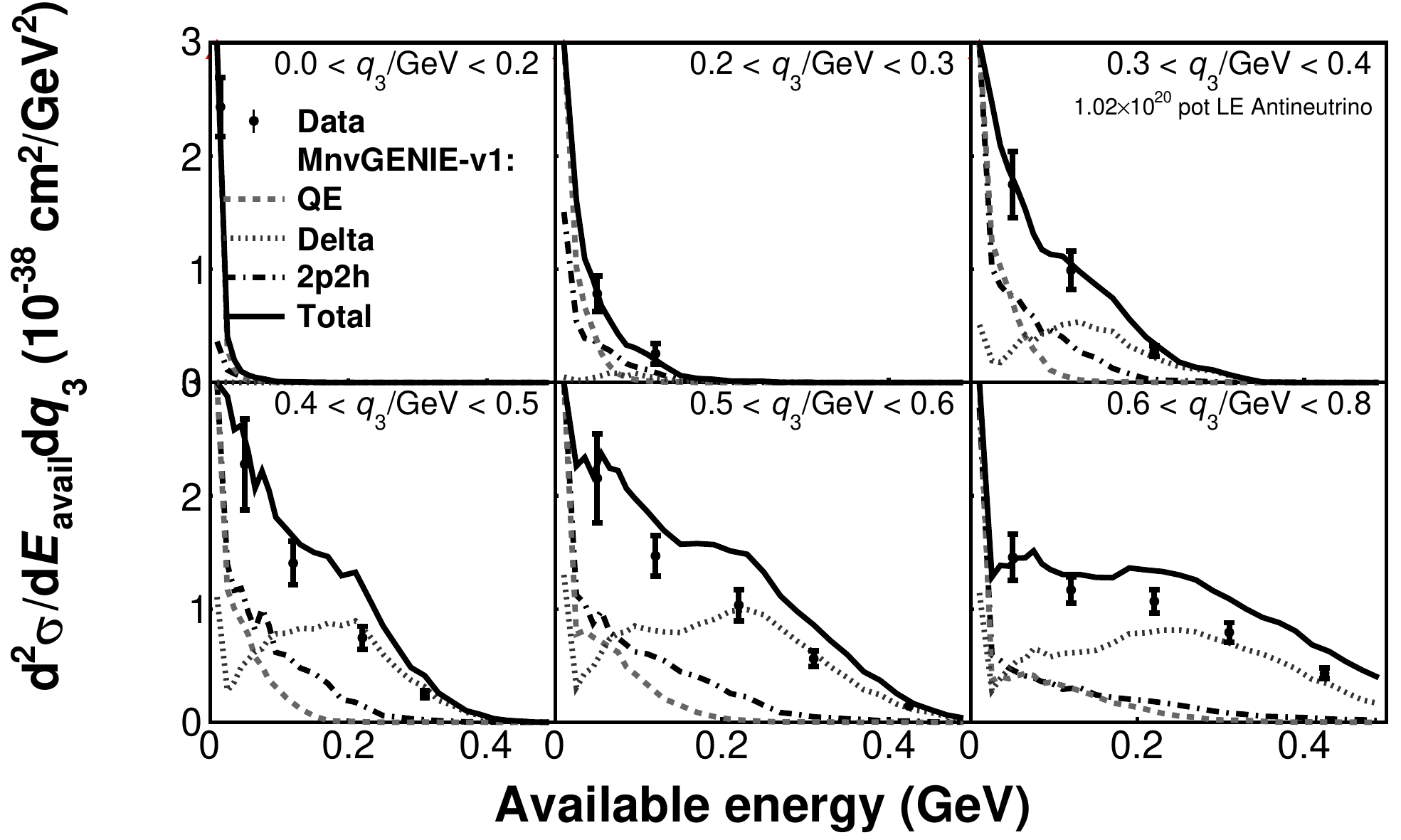}
\caption{Unfolded \dsdEdq cross section per nucleon compared to the model with RPA
  and tuned 2p2h components.  The breakdown of the predicted QE
  (dashed), 2p2h (dot-dashed),
  and $\Delta$ resonance (dotted) portions are shown.   To show the detail, 
the data and model prediction for the first bin (dominated by
neutron-only final states) are not shown;  they are far off the top
  of the plot with values between 9 and 17$\times 10^{-42}
  cm^{-2}/GeV^{2}$ per nucleon. 
\label{fig:unfolded}}
\end{center}
\end{figure}

To allow the development and testing of improved models, this distribution
is unfolded to produce a double differential cross section \dsdEdq,
shown in Fig.~\ref{fig:unfolded} and tabulated in the Supplemental Material. 
The procedure is the same as in \cite{Patrick:2018gvi}s, Sec. VIIB and VIII, 
and uses \cite{D'Agostini:1994zf,
  DAgostini:2010xxxxx, Adye:2011gm} but with three iterations.
The resolution for $q_3$ in Fig.~\ref{fig:resolution} is
with a rms near 23\% throughout and slowly changing
with $q_0$.
The reconstructed available energy is the sum of a component
from charged hadron and electromagnetic energy deposits with a central peak of
30\% resolution but rms of 40\% as in the neutrino case
\cite{Rodrigues:2015hik}.  Then the random tens of MeV energy from
about half of the final state neutrons further degrade the resolution
to Fig.~\ref{fig:resolution}.  
The 25 \Eavail , $q_3$ bins were chosen based on these resolutions.

The largest fractional uncertainties in half the bins, up to 14\%, come from
variations on the 2p2h enhancement used in the unfolding model.
When the enhancement is formed
only from events with {\it pn} initial state pairs (preferentially
{\it nn} final
states in the antineutrino case), the migration matrix has a higher
probability to put events in the low \Eavail bins.  The opposite is
true when the enhancement only adds {\it pp} initial state pairs. The difference to
the nominal cross section is added to the uncertainty.  The
uncertainty assigned to GENIE's intranuclear rescattering model is
also large because it modifies the unfolding model in this steep
region of the cross section.   These uncertainties are of similar size to
the flux uncertainty, suggesting that a future cycle of cross section model
improvements could  yield an even more precise cross section.
The breakdown of
uncertainties and the full covariance matrix 
are presented in the Supplemental Material.

In conclusion, the hadronic energy spectrum from a sample of
low momentum transfer antineutrino interactions suggests the need for a
RPA-like suppression \cite{Nieves:2004wx} of quasielastic events, relative to a Fermi
gas model.  In addition, an enhancement on top of the IFIC Valencia 2p2h
component \cite{Nieves:2011pp,Gran:2013kda} is essential to supply
the observed event rate in the region between the CCQE and $\Delta$ peaks.
We add to the evidence for a low $Q^2$ suppression of resonance events
by demonstrating that the MINOS parameterization \cite{Adamson:2014pgc}
offers some improvement to the $\chi^2$.
The model elements above were tested or fit to describe lepton and
hadronic components of neutrino data.
Critical for oscillation experiments in this neutrino energy range, they
offer a similarly good description of these antineutrino data.

  \begin{acknowledgments}
This docucment was prepared by members of the MINERvA collaboration 
using the resources of the Fermi National Accelerator Laboratory
(Fermilab), a U.S. Departent of Energy, Office of Science, HEP User
Facility.
Fermilab is managed by Fermi Research Alliance, LLC (FRA), acting
under Contract No. DE-AC02-07CH11359.   These resources included
support for the MINERvA construction project, and support for
construction was also granted by the United States National Science
Foundation under award PHY-0619727 and by the University of
Rochester.   Support for participating scientists was provided by NSF
and DOE (USA) by CAPES and CNPq (Brazil), by CoNaCyT (Mexico), by
Proyecto Basal FB 0821, CONICYT PIA ACT1413, Fondecyt 3170845 and
11130133 (Chile), by CONCYTEC, DGI-PUCP, and UDI/IGI-UNI (Peru), and
by the Latin American Center for Physics (CLAF).   We thank the MINOS
Collabortion for use of its near detector data.   Finally, we thank
the staff of Fermilab for support of the beamline, the detector, and
computing infrastructure.


\end{acknowledgments}

  \bibliographystyle{apsrev4-1}
  \bibliography{anuLowRecoil}

\begin{thebibliography}{60}%
\makeatletter
\providecommand \@ifxundefined [1]{%
 \@ifx{#1\undefined}
}%
\providecommand \@ifnum [1]{%
 \ifnum #1\expandafter \@firstoftwo
 \else \expandafter \@secondoftwo
 \fi
}%
\providecommand \@ifx [1]{%
 \ifx #1\expandafter \@firstoftwo
 \else \expandafter \@secondoftwo
 \fi
}%
\providecommand \natexlab [1]{#1}%
\providecommand \enquote  [1]{``#1''}%
\providecommand \bibnamefont  [1]{#1}%
\providecommand \bibfnamefont [1]{#1}%
\providecommand \citenamefont [1]{#1}%
\providecommand \href@noop [0]{\@secondoftwo}%
\providecommand \href [0]{\begingroup \@sanitize@url \@href}%
\providecommand \@href[1]{\@@startlink{#1}\@@href}%
\providecommand \@@href[1]{\endgroup#1\@@endlink}%
\providecommand \@sanitize@url [0]{\catcode `\\12\catcode `\$12\catcode
  `\&12\catcode `\#12\catcode `\^12\catcode `\_12\catcode `\%12\relax}%
\providecommand \@@startlink[1]{}%
\providecommand \@@endlink[0]{}%
\providecommand \url  [0]{\begingroup\@sanitize@url \@url }%
\providecommand \@url [1]{\endgroup\@href {#1}{\urlprefix }}%
\providecommand \urlprefix  [0]{URL }%
\providecommand \Eprint [0]{\href }%
\providecommand \doibase [0]{http://dx.doi.org/}%
\providecommand \selectlanguage [0]{\@gobble}%
\providecommand \bibinfo  [0]{\@secondoftwo}%
\providecommand \bibfield  [0]{\@secondoftwo}%
\providecommand \translation [1]{[#1]}%
\providecommand \BibitemOpen [0]{}%
\providecommand \bibitemStop [0]{}%
\providecommand \bibitemNoStop [0]{.\EOS\space}%
\providecommand \EOS [0]{\spacefactor3000\relax}%
\providecommand \BibitemShut  [1]{\csname bibitem#1\endcsname}%
\let\auto@bib@innerbib\@empty
\bibitem [{\citenamefont {Adamson}\ \emph
  {et~al.}(2016{\natexlab{a}})\citenamefont {Adamson} \emph
  {et~al.}}]{Adamson:2016xxw}%
  \BibitemOpen
  \bibfield  {author} {\bibinfo {author} {\bibfnamefont {P.}~\bibnamefont
  {Adamson}} \emph {et~al.} (\bibinfo {collaboration} {NOvA Collaboration}),\
  }\href {\doibase 10.1103/PhysRevD.93.051104} {\bibfield  {journal} {\bibinfo
  {journal} {Phys.~Rev.~D}\ }\textbf {\bibinfo {volume} {93}},\ \bibinfo
  {pages} {051104} (\bibinfo {year} {2016}{\natexlab{a}})},\ \Eprint
  {http://arxiv.org/abs/1601.05037} {arXiv:1601.05037} \BibitemShut {NoStop}%
\bibitem [{\citenamefont {Adamson}\ \emph
  {et~al.}(2016{\natexlab{b}})\citenamefont {Adamson} \emph
  {et~al.}}]{Adamson:2016tbq}%
  \BibitemOpen
  \bibfield  {author} {\bibinfo {author} {\bibfnamefont {P.}~\bibnamefont
  {Adamson}} \emph {et~al.} (\bibinfo {collaboration} {NOvA Collaboration}),\
  }\href {\doibase 10.1103/PhysRevLett.116.151806} {\bibfield  {journal}
  {\bibinfo  {journal} {Phys. Rev. Lett.}\ }\textbf {\bibinfo {volume} {116}},\
  \bibinfo {pages} {151806} (\bibinfo {year} {2016}{\natexlab{b}})},\ \Eprint
  {http://arxiv.org/abs/1601.05022} {arXiv:1601.05022} \BibitemShut {NoStop}%
\bibitem [{\citenamefont {Acciarri}\ \emph
  {et~al.}(2015{\natexlab{a}})\citenamefont {Acciarri} \emph
  {et~al.}}]{Acciari:2015zzz}%
  \BibitemOpen
  \bibfield  {author} {\bibinfo {author} {\bibfnamefont {R.}~\bibnamefont
  {Acciarri}} \emph {et~al.} (\bibinfo {collaboration} {Fermilab Short Baseline
  Collaboration}),\ }\href@noop {} {\  (\bibinfo {year}
  {2015}{\natexlab{a}})},\ \Eprint {http://arxiv.org/abs/1503.01520}
  {arXiv:1503.01520} \BibitemShut {NoStop}%
\bibitem [{\citenamefont {Adams}\ \emph {et~al.}(2013)\citenamefont {Adams}
  \emph {et~al.}}]{Adams:2013qkq}%
  \BibitemOpen
  \bibfield  {author} {\bibinfo {author} {\bibfnamefont {C.}~\bibnamefont
  {Adams}} \emph {et~al.} (\bibinfo {collaboration} {LBNE Collaboration})\
  }(\bibinfo {year} {2013})\ \Eprint {http://arxiv.org/abs/1307.7335}
  {arXiv:1307.7335} \BibitemShut {NoStop}%
\bibitem [{\citenamefont {Acciarri}\ \emph
  {et~al.}(2015{\natexlab{b}})\citenamefont {Acciarri} \emph
  {et~al.}}]{Acciarri:2015uup}%
  \BibitemOpen
  \bibfield  {author} {\bibinfo {author} {\bibfnamefont {R.}~\bibnamefont
  {Acciarri}} \emph {et~al.} (\bibinfo {collaboration} {DUNE Collaboration}),\
  }\href@noop {} {\  (\bibinfo {year} {2015}{\natexlab{b}})},\ \Eprint
  {http://arxiv.org/abs/1512.06148} {arXiv:1512.06148} \BibitemShut {NoStop}%
\bibitem [{\citenamefont {Rodrigues}\ \emph
  {et~al.}(2016{\natexlab{a}})\citenamefont {Rodrigues} \emph
  {et~al.}}]{Rodrigues:2015hik}%
  \BibitemOpen
  \bibfield  {author} {\bibinfo {author} {\bibfnamefont {P.~A.}\ \bibnamefont
  {Rodrigues}} \emph {et~al.} (\bibinfo {collaboration} {MINERvA
  Collaboration}),\ }\href {\doibase 10.1103/PhysRevLett.116.071802} {\bibfield
   {journal} {\bibinfo  {journal} {Phys. Rev. Lett.}\ }\textbf {\bibinfo
  {volume} {116}},\ \bibinfo {pages} {071802} (\bibinfo {year}
  {2016}{\natexlab{a}})},\ \Eprint {http://arxiv.org/abs/1511.05944}
  {arXiv:1511.05944} \BibitemShut {NoStop}%
\bibitem [{\citenamefont {Martini}\ \emph {et~al.}(2009)\citenamefont
  {Martini}, \citenamefont {Ericson}, \citenamefont {Chanfray},\ and\
  \citenamefont {Marteau}}]{Martini:2009uj}%
  \BibitemOpen
  \bibfield  {author} {\bibinfo {author} {\bibfnamefont {M.}~\bibnamefont
  {Martini}}, \bibinfo {author} {\bibfnamefont {M.}~\bibnamefont {Ericson}},
  \bibinfo {author} {\bibfnamefont {G.}~\bibnamefont {Chanfray}}, \ and\
  \bibinfo {author} {\bibfnamefont {J.}~\bibnamefont {Marteau}},\ }\href
  {\doibase 10.1103/PhysRevC.80.065501} {\bibfield  {journal} {\bibinfo
  {journal} {Phys.~Rev.~C}\ }\textbf {\bibinfo {volume} {80}},\ \bibinfo
  {pages} {065501} (\bibinfo {year} {2009})},\ \Eprint
  {http://arxiv.org/abs/0910.2622} {arXiv:0910.2622} \BibitemShut {NoStop}%
\bibitem [{\citenamefont {Nieves}\ \emph {et~al.}(2011)\citenamefont {Nieves},
  \citenamefont {Ruiz~Simo},\ and\ \citenamefont
  {Vicente~Vacas}}]{Nieves:2011pp}%
  \BibitemOpen
  \bibfield  {author} {\bibinfo {author} {\bibfnamefont {J.}~\bibnamefont
  {Nieves}}, \bibinfo {author} {\bibfnamefont {I.}~\bibnamefont {Ruiz~Simo}}, \
  and\ \bibinfo {author} {\bibfnamefont {M.}~\bibnamefont {Vicente~Vacas}},\
  }\href {\doibase 10.1103/PhysRevC.83.045501} {\bibfield  {journal} {\bibinfo
  {journal} {Phys.~Rev.~C}\ }\textbf {\bibinfo {volume} {83}},\ \bibinfo
  {pages} {045501} (\bibinfo {year} {2011})},\ \Eprint
  {http://arxiv.org/abs/1102.2777} {arXiv:1102.2777} \BibitemShut {NoStop}%
\bibitem [{\citenamefont {Gonzaléz-Jiménez}\ \emph
  {et~al.}(2014)\citenamefont {Gonzaléz-Jiménez}, \citenamefont {Megias},
  \citenamefont {Barbaro}, \citenamefont {Caballero},\ and\ \citenamefont
  {Donnelly}}]{Gonzalez-Jimenez:2014eqa}%
  \BibitemOpen
  \bibfield  {author} {\bibinfo {author} {\bibfnamefont {R.}~\bibnamefont
  {Gonzaléz-Jiménez}}, \bibinfo {author} {\bibfnamefont {G.~D.}\ \bibnamefont
  {Megias}}, \bibinfo {author} {\bibfnamefont {M.~B.}\ \bibnamefont {Barbaro}},
  \bibinfo {author} {\bibfnamefont {J.~A.}\ \bibnamefont {Caballero}}, \ and\
  \bibinfo {author} {\bibfnamefont {T.~W.}\ \bibnamefont {Donnelly}},\ }\href
  {\doibase 10.1103/PhysRevC.90.035501} {\bibfield  {journal} {\bibinfo
  {journal} {Phys.~Rev.~C}\ }\textbf {\bibinfo {volume} {90}},\ \bibinfo
  {pages} {035501} (\bibinfo {year} {2014})},\ \Eprint
  {http://arxiv.org/abs/1407.8346} {arXiv:1407.8346} \BibitemShut {NoStop}%
\bibitem [{\citenamefont {Megias}\ \emph {et~al.}(2016)\citenamefont {Megias},
  \citenamefont {Amaro}, \citenamefont {Barbaro}, \citenamefont {Caballero},
  \citenamefont {Donnelly},\ and\ \citenamefont {Ruiz~Simo}}]{Megias:2016fjk}%
  \BibitemOpen
  \bibfield  {author} {\bibinfo {author} {\bibfnamefont {G.}~\bibnamefont
  {Megias}}, \bibinfo {author} {\bibfnamefont {J.}~\bibnamefont {Amaro}},
  \bibinfo {author} {\bibfnamefont {M.}~\bibnamefont {Barbaro}}, \bibinfo
  {author} {\bibfnamefont {J.}~\bibnamefont {Caballero}}, \bibinfo {author}
  {\bibfnamefont {T.}~\bibnamefont {Donnelly}}, \ and\ \bibinfo {author}
  {\bibfnamefont {I.}~\bibnamefont {Ruiz~Simo}},\ }\href {\doibase
  10.1103/PhysRevD.94.093004} {\bibfield  {journal} {\bibinfo  {journal}
  {Phys.~Rev.~D}\ }\textbf {\bibinfo {volume} {94}},\ \bibinfo {pages} {093004}
  (\bibinfo {year} {2016})},\ \Eprint {http://arxiv.org/abs/1607.08565}
  {arXiv:1607.08565} \BibitemShut {NoStop}%
\bibitem [{\citenamefont {Van~Cuyck}\ \emph {et~al.}(2017)\citenamefont
  {Van~Cuyck}, \citenamefont {Jachowicz}, \citenamefont {González-Jiménez},
  \citenamefont {Ryckebusch},\ and\ \citenamefont
  {Van~Dessel}}]{VanCuyck:2017wfn}%
  \BibitemOpen
  \bibfield  {author} {\bibinfo {author} {\bibfnamefont {T.}~\bibnamefont
  {Van~Cuyck}}, \bibinfo {author} {\bibfnamefont {N.}~\bibnamefont
  {Jachowicz}}, \bibinfo {author} {\bibfnamefont {R.}~\bibnamefont
  {González-Jiménez}}, \bibinfo {author} {\bibfnamefont {J.}~\bibnamefont
  {Ryckebusch}}, \ and\ \bibinfo {author} {\bibfnamefont {N.}~\bibnamefont
  {Van~Dessel}},\ }\href {\doibase 10.1103/PhysRevC.95.054611} {\bibfield
  {journal} {\bibinfo  {journal} {Phys.~Rev.~C}\ }\textbf {\bibinfo {volume}
  {95}},\ \bibinfo {pages} {054611} (\bibinfo {year} {2017})},\ \Eprint
  {http://arxiv.org/abs/1702.06402} {arXiv:1702.06402} \BibitemShut {NoStop}%
\bibitem [{\citenamefont {Ruiz~Simo}\ \emph {et~al.}(2018)\citenamefont
  {Ruiz~Simo}, \citenamefont {Amaro}, \citenamefont {Barbaro}, \citenamefont
  {Caballero}, \citenamefont {Megias},\ and\ \citenamefont
  {Donnelly}}]{RuizSimo:2017hlc}%
  \BibitemOpen
  \bibfield  {author} {\bibinfo {author} {\bibfnamefont {I.}~\bibnamefont
  {Ruiz~Simo}}, \bibinfo {author} {\bibfnamefont {J.~E.}\ \bibnamefont
  {Amaro}}, \bibinfo {author} {\bibfnamefont {M.~B.}\ \bibnamefont {Barbaro}},
  \bibinfo {author} {\bibfnamefont {J.~A.}\ \bibnamefont {Caballero}}, \bibinfo
  {author} {\bibfnamefont {G.~D.}\ \bibnamefont {Megias}}, \ and\ \bibinfo
  {author} {\bibfnamefont {T.~W.}\ \bibnamefont {Donnelly}},\ }\href {\doibase
  10.1016/j.aop.2017.11.029} {\bibfield  {journal} {\bibinfo  {journal} {Annals
  Phys.}\ }\textbf {\bibinfo {volume} {388}},\ \bibinfo {pages} {323} (\bibinfo
  {year} {2018})},\ \Eprint {http://arxiv.org/abs/1706.06377}
  {arXiv:1706.06377} \BibitemShut {NoStop}%
\bibitem [{\citenamefont {Lightbody}\ and\ \citenamefont
  {O'Connell}(1988)}]{Lightbody:1988}%
  \BibitemOpen
  \bibfield  {author} {\bibinfo {author} {\bibfnamefont {J.~W.}\ \bibnamefont
  {Lightbody}}\ and\ \bibinfo {author} {\bibfnamefont {J.~S.}\ \bibnamefont
  {O'Connell}},\ }\href@noop {} {\bibfield  {journal} {\bibinfo  {journal}
  {Computers in Physics}\ }\textbf {\bibinfo {volume} {2}},\ \bibinfo {pages}
  {57} (\bibinfo {year} {1988})}\BibitemShut {NoStop}%
\bibitem [{\citenamefont {Sobczyk}(2012)}]{Sobczyk:2012ms}%
  \BibitemOpen
  \bibfield  {author} {\bibinfo {author} {\bibfnamefont {J.~T.}\ \bibnamefont
  {Sobczyk}},\ }\href {\doibase 10.1103/PhysRevC.86.015504} {\bibfield
  {journal} {\bibinfo  {journal} {Phys.~Rev.~C}\ }\textbf {\bibinfo {volume}
  {86}},\ \bibinfo {pages} {015504} (\bibinfo {year} {2012})},\ \Eprint
  {http://arxiv.org/abs/1201.3673} {arXiv:1201.3673} \BibitemShut {NoStop}%
\bibitem [{\citenamefont {Schwehr}\ \emph {et~al.}(2016)\citenamefont
  {Schwehr}, \citenamefont {Cherdack},\ and\ \citenamefont
  {Gran}}]{Schwehr:2016pvn}%
  \BibitemOpen
  \bibfield  {author} {\bibinfo {author} {\bibfnamefont {J.}~\bibnamefont
  {Schwehr}}, \bibinfo {author} {\bibfnamefont {D.}~\bibnamefont {Cherdack}}, \
  and\ \bibinfo {author} {\bibfnamefont {R.}~\bibnamefont {Gran}},\ }\href@noop
  {} {\  (\bibinfo {year} {2016})},\ \Eprint {http://arxiv.org/abs/1601.02038}
  {arXiv:1601.02038} \BibitemShut {NoStop}%
\bibitem [{\citenamefont {Gallmeister}\ \emph {et~al.}(2016)\citenamefont
  {Gallmeister}, \citenamefont {Mosel},\ and\ \citenamefont
  {Weil}}]{Gallmeister:2016dnq}%
  \BibitemOpen
  \bibfield  {author} {\bibinfo {author} {\bibfnamefont {K.}~\bibnamefont
  {Gallmeister}}, \bibinfo {author} {\bibfnamefont {U.}~\bibnamefont {Mosel}},
  \ and\ \bibinfo {author} {\bibfnamefont {J.}~\bibnamefont {Weil}},\ }\href
  {\doibase 10.1103/PhysRevC.94.035502} {\bibfield  {journal} {\bibinfo
  {journal} {Phys.~Rev.~C}\ }\textbf {\bibinfo {volume} {94}},\ \bibinfo
  {pages} {035502} (\bibinfo {year} {2016})},\ \Eprint
  {http://arxiv.org/abs/1605.09391} {arXiv:1605.09391} \BibitemShut {NoStop}%
\bibitem [{\citenamefont {Nieves}\ \emph {et~al.}(2004)\citenamefont {Nieves},
  \citenamefont {Amaro},\ and\ \citenamefont {Valverde}}]{Nieves:2004wx}%
  \BibitemOpen
  \bibfield  {author} {\bibinfo {author} {\bibfnamefont {J.}~\bibnamefont
  {Nieves}}, \bibinfo {author} {\bibfnamefont {J.~E.}\ \bibnamefont {Amaro}}, \
  and\ \bibinfo {author} {\bibfnamefont {M.}~\bibnamefont {Valverde}},\ }\href
  {\doibase 10.1103/PhysRevC.70.055503, 10.1103/PhysRevC.72.019902} {\bibfield
  {journal} {\bibinfo  {journal} {Phys.~Rev.~C}\ }\textbf {\bibinfo {volume}
  {70}},\ \bibinfo {pages} {055503} (\bibinfo {year} {2004})},\ \Eprint
  {http://arxiv.org/abs/nucl-th/0408005} {arXiv:nucl-th/0408005} \BibitemShut
  {NoStop}%
\bibitem [{\citenamefont {Pandey}\ \emph {et~al.}(2015)\citenamefont {Pandey},
  \citenamefont {Jachowicz}, \citenamefont {Van~Cuyck}, \citenamefont
  {Ryckebusch},\ and\ \citenamefont {Martini}}]{Pandey:2014tza}%
  \BibitemOpen
  \bibfield  {author} {\bibinfo {author} {\bibfnamefont {V.}~\bibnamefont
  {Pandey}}, \bibinfo {author} {\bibfnamefont {N.}~\bibnamefont {Jachowicz}},
  \bibinfo {author} {\bibfnamefont {T.}~\bibnamefont {Van~Cuyck}}, \bibinfo
  {author} {\bibfnamefont {J.}~\bibnamefont {Ryckebusch}}, \ and\ \bibinfo
  {author} {\bibfnamefont {M.}~\bibnamefont {Martini}},\ }\href {\doibase
  10.1103/PhysRevC.92.024606} {\bibfield  {journal} {\bibinfo  {journal}
  {Phys.~Rev.~C}\ }\textbf {\bibinfo {volume} {92}},\ \bibinfo {pages} {024606}
  (\bibinfo {year} {2015})},\ \Eprint {http://arxiv.org/abs/1412.4624}
  {arXiv:1412.4624} \BibitemShut {NoStop}%
\bibitem [{\citenamefont {Nieves}\ and\ \citenamefont
  {Sobczyk}(2017)}]{Nieves:2017lij}%
  \BibitemOpen
  \bibfield  {author} {\bibinfo {author} {\bibfnamefont {J.}~\bibnamefont
  {Nieves}}\ and\ \bibinfo {author} {\bibfnamefont {J.~E.}\ \bibnamefont
  {Sobczyk}},\ }\href {\doibase 10.1016/j.aop.2017.06.002} {\bibfield
  {journal} {\bibinfo  {journal} {Annals Phys.}\ }\textbf {\bibinfo {volume}
  {383}},\ \bibinfo {pages} {455} (\bibinfo {year} {2017})},\ \Eprint
  {http://arxiv.org/abs/1701.03628} {arXiv:1701.03628} \BibitemShut {NoStop}%
\bibitem [{\citenamefont {Smith}\ and\ \citenamefont
  {Moniz}(1972)}]{Smith:1972xh}%
  \BibitemOpen
  \bibfield  {author} {\bibinfo {author} {\bibfnamefont {R.~A.}\ \bibnamefont
  {Smith}}\ and\ \bibinfo {author} {\bibfnamefont {E.~J.}\ \bibnamefont
  {Moniz}},\ }\href {\doibase 10.1016/0550-3213(72)90040-5} {\bibfield
  {journal} {\bibinfo  {journal} {Nucl.~Phys.~B}\ }\textbf {\bibinfo {volume}
  {43}},\ \bibinfo {pages} {605} (\bibinfo {year} {1972})}\BibitemShut
  {NoStop}%
\bibitem [{\citenamefont {Adamson}\ \emph
  {et~al.}(2016{\natexlab{c}})\citenamefont {Adamson} \emph
  {et~al.}}]{Adamson:2015dkw}%
  \BibitemOpen
  \bibfield  {author} {\bibinfo {author} {\bibfnamefont {P.}~\bibnamefont
  {Adamson}} \emph {et~al.},\ }\href {\doibase 10.1016/j.nima.2015.08.063}
  {\bibfield  {journal} {\bibinfo  {journal} {Nucl.~Instrum.~Meth.~A}\ }\textbf
  {\bibinfo {volume} {806}},\ \bibinfo {pages} {279} (\bibinfo {year}
  {2016}{\natexlab{c}})},\ \Eprint {http://arxiv.org/abs/1507.06690}
  {arXiv:1507.06690} \BibitemShut {NoStop}%
\bibitem [{\citenamefont {Agostinelli}\ \emph {et~al.}(2003)\citenamefont
  {Agostinelli} \emph {et~al.}}]{Agostinelli2003250}%
  \BibitemOpen
  \bibfield  {author} {\bibinfo {author} {\bibfnamefont {S.}~\bibnamefont
  {Agostinelli}} \emph {et~al.},\ }\href {\doibase
  10.1016/S0168-9002(03)01368-8} {\bibfield  {journal} {\bibinfo  {journal}
  {Nucl.~Instrum.~Meth.~A}\ }\textbf {\bibinfo {volume} {506}},\ \bibinfo
  {pages} {250 } (\bibinfo {year} {2003})}\BibitemShut {NoStop}%
\bibitem [{\citenamefont {Allison}\ \emph {et~al.}(2006)\citenamefont {Allison}
  \emph {et~al.}}]{1610988}%
  \BibitemOpen
  \bibfield  {author} {\bibinfo {author} {\bibfnamefont {J.}~\bibnamefont
  {Allison}} \emph {et~al.},\ }\href {\doibase 10.1109/TNS.2006.869826}
  {\bibfield  {journal} {\bibinfo  {journal} {Nuclear Science, IEEE
  Transactions on}\ }\textbf {\bibinfo {volume} {53}},\ \bibinfo {pages} {270}
  (\bibinfo {year} {2006})}\BibitemShut {NoStop}%
\bibitem [{\citenamefont {Aliaga}\ \emph {et~al.}(2016)\citenamefont {Aliaga}
  \emph {et~al.}}]{Aliaga:2016oaz}%
  \BibitemOpen
  \bibfield  {author} {\bibinfo {author} {\bibfnamefont {L.}~\bibnamefont
  {Aliaga}} \emph {et~al.} (\bibinfo {collaboration} {MINERvA Collaboration}),\
  }\href {\doibase 10.1103/PhysRevD.94.092005, 10.1103/PhysRevD.95.039903}
  {\bibfield  {journal} {\bibinfo  {journal} {Phys.~Rev.~D}\ }\textbf {\bibinfo
  {volume} {94}},\ \bibinfo {pages} {092005} (\bibinfo {year} {2016})},\
  \bibinfo {note} {[Addendum: Phys. Rev.D 95,no.3,039903(2017)]},\ \Eprint
  {http://arxiv.org/abs/1607.00704} {arXiv:1607.00704} \BibitemShut {NoStop}%
\bibitem [{\citenamefont {Alt}\ \emph {et~al.}(2007)\citenamefont {Alt} \emph
  {et~al.}}]{Alt:2006fr}%
  \BibitemOpen
  \bibfield  {author} {\bibinfo {author} {\bibfnamefont {C.}~\bibnamefont
  {Alt}} \emph {et~al.} (\bibinfo {collaboration} {NA49 Collaboration}),\
  }\href {\doibase 10.1140/epjc/s10052-006-0165-7} {\bibfield  {journal}
  {\bibinfo  {journal} {Eur.~Phys.~J.~C}\ }\textbf {\bibinfo {volume} {49}},\
  \bibinfo {pages} {897} (\bibinfo {year} {2007})},\ \Eprint
  {http://arxiv.org/abs/hep-ex/0606028} {arXiv:hep-ex/0606028} \BibitemShut
  {NoStop}%
\bibitem [{\citenamefont {Denisov}\ \emph {et~al.}(1973)\citenamefont
  {Denisov}, \citenamefont {Donskov}, \citenamefont {Gorin}, \citenamefont
  {Krasnokutsky}, \citenamefont {Petrukhin}, \citenamefont {Prokoshkin},\ and\
  \citenamefont {Stoyanova}}]{Denisov:1973zv}%
  \BibitemOpen
  \bibfield  {author} {\bibinfo {author} {\bibfnamefont {S.~P.}\ \bibnamefont
  {Denisov}}, \bibinfo {author} {\bibfnamefont {S.~V.}\ \bibnamefont
  {Donskov}}, \bibinfo {author} {\bibfnamefont {{\relax Yu}.~P.}\ \bibnamefont
  {Gorin}}, \bibinfo {author} {\bibfnamefont {R.~N.}\ \bibnamefont
  {Krasnokutsky}}, \bibinfo {author} {\bibfnamefont {A.~I.}\ \bibnamefont
  {Petrukhin}}, \bibinfo {author} {\bibfnamefont {{\relax Yu}.~D.}\
  \bibnamefont {Prokoshkin}}, \ and\ \bibinfo {author} {\bibfnamefont {D.~A.}\
  \bibnamefont {Stoyanova}},\ }\href {\doibase 10.1016/0550-3213(73)90351-9}
  {\bibfield  {journal} {\bibinfo  {journal} {Nucl.~Phys.~B}\ }\textbf
  {\bibinfo {volume} {61}},\ \bibinfo {pages} {62} (\bibinfo {year}
  {1973})}\BibitemShut {NoStop}%
\bibitem [{\citenamefont {Carroll}\ \emph {et~al.}(1979)\citenamefont {Carroll}
  \emph {et~al.}}]{Carroll:1978hc}%
  \BibitemOpen
  \bibfield  {author} {\bibinfo {author} {\bibfnamefont {A.~S.}\ \bibnamefont
  {Carroll}} \emph {et~al.},\ }\href {\doibase 10.1016/0370-2693(79)90226-0}
  {\bibfield  {journal} {\bibinfo  {journal} {Phys.~Lett.~B}\ }\textbf
  {\bibinfo {volume} {80}},\ \bibinfo {pages} {319} (\bibinfo {year}
  {1979})}\BibitemShut {NoStop}%
\bibitem [{\citenamefont {Allaby}\ \emph {et~al.}(1969)\citenamefont {Allaby}
  \emph {et~al.}}]{Allaby:1969de}%
  \BibitemOpen
  \bibfield  {author} {\bibinfo {author} {\bibfnamefont {J.~V.}\ \bibnamefont
  {Allaby}} \emph {et~al.} (\bibinfo {collaboration} {IHEP-CERN
  Collaboration}),\ }\href {\doibase 10.1016/0370-2693(69)90184-1} {\bibfield
  {journal} {\bibinfo  {journal} {Phys.~Lett.~B}\ }\textbf {\bibinfo {volume}
  {30}},\ \bibinfo {pages} {500} (\bibinfo {year} {1969})}\BibitemShut
  {NoStop}%
\bibitem [{\citenamefont {Park}\ \emph {et~al.}(2016)\citenamefont {Park} \emph
  {et~al.}}]{Park:2015eqa}%
  \BibitemOpen
  \bibfield  {author} {\bibinfo {author} {\bibfnamefont {J.}~\bibnamefont
  {Park}} \emph {et~al.} (\bibinfo {collaboration} {MINERvA Collaboration}),\
  }\href {\doibase 10.1103/PhysRevD.93.112007} {\bibfield  {journal} {\bibinfo
  {journal} {Phys.~Rev.~D}\ }\textbf {\bibinfo {volume} {93}},\ \bibinfo
  {pages} {112007} (\bibinfo {year} {2016})},\ \Eprint
  {http://arxiv.org/abs/1512.07699} {arXiv:1512.07699} \BibitemShut {NoStop}%
\bibitem [{\citenamefont {Michael}\ \emph {et~al.}(2008)\citenamefont {Michael}
  \emph {et~al.}}]{Michael:2008bc}%
  \BibitemOpen
  \bibfield  {author} {\bibinfo {author} {\bibfnamefont {D.~G.}\ \bibnamefont
  {Michael}} \emph {et~al.} (\bibinfo {collaboration} {MINOS Collaboration}),\
  }\href {\doibase 10.1016/j.nima.2008.08.003} {\bibfield  {journal} {\bibinfo
  {journal} {Nucl.~Instrum.~Meth.~A}\ }\textbf {\bibinfo {volume} {596}},\
  \bibinfo {pages} {190} (\bibinfo {year} {2008})},\ \Eprint
  {http://arxiv.org/abs/0805.3170} {arXiv:0805.3170} \BibitemShut {NoStop}%
\bibitem [{\citenamefont {Aliaga}\ \emph {et~al.}(2014)\citenamefont {Aliaga}
  \emph {et~al.}}]{Aliaga:2013uqz}%
  \BibitemOpen
  \bibfield  {author} {\bibinfo {author} {\bibfnamefont {L.}~\bibnamefont
  {Aliaga}} \emph {et~al.} (\bibinfo {collaboration} {MINERvA Collaboration}),\
  }\href {\doibase 10.1016/j.nima.2013.12.053} {\bibfield  {journal} {\bibinfo
  {journal} {Nucl.~Instrum.~Meth. A}\ }\textbf {\bibinfo {volume} {743}},\
  \bibinfo {pages} {130} (\bibinfo {year} {2014})},\ \Eprint
  {http://arxiv.org/abs/1305.5199} {arXiv:1305.5199} \BibitemShut {NoStop}%
\bibitem [{\citenamefont {Aliaga}\ \emph {et~al.}(2015)\citenamefont {Aliaga}
  \emph {et~al.}}]{Aliaga:2015aqe}%
  \BibitemOpen
  \bibfield  {author} {\bibinfo {author} {\bibfnamefont {L.}~\bibnamefont
  {Aliaga}} \emph {et~al.} (\bibinfo {collaboration} {MINERvA Collaboration}),\
  }\href {\doibase 10.1016/j.nima.2015.04.003} {\bibfield  {journal} {\bibinfo
  {journal} {Nucl.~Instrum.~Meth.~A}\ }\textbf {\bibinfo {volume} {789}},\
  \bibinfo {pages} {28} (\bibinfo {year} {2015})},\ \Eprint
  {http://arxiv.org/abs/1501.06431} {arXiv:1501.06431} \BibitemShut {NoStop}%
\bibitem [{\citenamefont {Abfalterer}\ \emph {et~al.}(2001)\citenamefont
  {Abfalterer}, \citenamefont {Bateman}, \citenamefont {Dietrich},
  \citenamefont {Finlay}, \citenamefont {Haight} \emph
  {et~al.}}]{Abfalterer:2001gw}%
  \BibitemOpen
  \bibfield  {author} {\bibinfo {author} {\bibfnamefont {W.~P.}\ \bibnamefont
  {Abfalterer}}, \bibinfo {author} {\bibfnamefont {F.~B.}\ \bibnamefont
  {Bateman}}, \bibinfo {author} {\bibfnamefont {F.~S.}\ \bibnamefont
  {Dietrich}}, \bibinfo {author} {\bibfnamefont {R.~W.}\ \bibnamefont
  {Finlay}}, \bibinfo {author} {\bibfnamefont {R.~C.}\ \bibnamefont {Haight}},
  \emph {et~al.},\ }\href {\doibase 10.1103/PhysRevC.63.044608} {\bibfield
  {journal} {\bibinfo  {journal} {Phys.~Rev.~C}\ }\textbf {\bibinfo {volume}
  {63}},\ \bibinfo {pages} {044608} (\bibinfo {year} {2001})}\BibitemShut
  {NoStop}%
\bibitem [{\citenamefont {Andreopoulos}\ \emph {et~al.}(2010)\citenamefont
  {Andreopoulos} \emph {et~al.}}]{Andreopoulos201087}%
  \BibitemOpen
  \bibfield  {author} {\bibinfo {author} {\bibfnamefont {C.}~\bibnamefont
  {Andreopoulos}} \emph {et~al.},\ }\href {\doibase 10.1016/j.nima.2009.12.009}
  {\bibfield  {journal} {\bibinfo  {journal} {Nucl.~Instrum.~Meth.~A}\ }\textbf
  {\bibinfo {volume} {614}},\ \bibinfo {pages} {87 } (\bibinfo {year}
  {2010})},\ \bibinfo {note} {{P}rogram version 2.8.4, with private
  modifications, used here}\BibitemShut {NoStop}%
\bibitem [{\citenamefont {Devan}(2015)}]{Devan:2015uak}%
  \BibitemOpen
  \bibfield  {author} {\bibinfo {author} {\bibfnamefont {J.~D.}\ \bibnamefont
  {Devan}},\ }\emph {\bibinfo {title} {{Measurement of Neutrino and
  Antineutrino Charged-Current Inclusive Cross Sections with the MINERvA
  Detector}}},\ \href {\doibase 10.2172/1248217} {Ph.D. thesis},\ \bibinfo
  {school} {Coll. William and Mary} (\bibinfo {year} {2015})\BibitemShut
  {NoStop}%
\bibitem [{\citenamefont {Llewellyn~Smith}(1972)}]{LlewellynSmith:1971zm}%
  \BibitemOpen
  \bibfield  {author} {\bibinfo {author} {\bibfnamefont {C.~H.}\ \bibnamefont
  {Llewellyn~Smith}},\ }\href {\doibase 10.1016/0370-1573(72)90010-5}
  {\bibfield  {journal} {\bibinfo  {journal} {Phys.~Rept.}\ }\textbf {\bibinfo
  {volume} {3}},\ \bibinfo {pages} {261} (\bibinfo {year} {1972})}\BibitemShut
  {NoStop}%
\bibitem [{\citenamefont {Bradford}\ \emph {et~al.}(2006)\citenamefont
  {Bradford}, \citenamefont {Bodek}, \citenamefont {Budd},\ and\ \citenamefont
  {Arrington}}]{Bradford:2006yz}%
  \BibitemOpen
  \bibfield  {author} {\bibinfo {author} {\bibfnamefont {R.}~\bibnamefont
  {Bradford}}, \bibinfo {author} {\bibfnamefont {A.}~\bibnamefont {Bodek}},
  \bibinfo {author} {\bibfnamefont {H.~S.}\ \bibnamefont {Budd}}, \ and\
  \bibinfo {author} {\bibfnamefont {J.}~\bibnamefont {Arrington}},\ }\href
  {\doibase 10.1016/j.nuclphysbps.2006.08.028} {\bibfield  {journal} {\bibinfo
  {journal} {Nucl.~Phys.~Proc.~Suppl.}\ }\textbf {\bibinfo {volume} {159}},\
  \bibinfo {pages} {127} (\bibinfo {year} {2006})},\ \Eprint
  {http://arxiv.org/abs/hep-ex/0602017} {arXiv:hep-ex/0602017} \BibitemShut
  {NoStop}%
\bibitem [{\citenamefont {Rein}\ and\ \citenamefont
  {Sehgal}(1981)}]{Rein:1980wg}%
  \BibitemOpen
  \bibfield  {author} {\bibinfo {author} {\bibfnamefont {D.}~\bibnamefont
  {Rein}}\ and\ \bibinfo {author} {\bibfnamefont {L.~M.}\ \bibnamefont
  {Sehgal}},\ }\href {\doibase 10.1016/0003-4916(81)90242-6} {\bibfield
  {journal} {\bibinfo  {journal} {Annals~Phys.}\ }\textbf {\bibinfo {volume}
  {133}},\ \bibinfo {pages} {79} (\bibinfo {year} {1981})}\BibitemShut
  {NoStop}%
\bibitem [{\citenamefont {Bodek}\ \emph {et~al.}(2005)\citenamefont {Bodek},
  \citenamefont {Park},\ and\ \citenamefont {Yang}}]{Bodek:2004pc}%
  \BibitemOpen
  \bibfield  {author} {\bibinfo {author} {\bibfnamefont {A.}~\bibnamefont
  {Bodek}}, \bibinfo {author} {\bibfnamefont {I.}~\bibnamefont {Park}}, \ and\
  \bibinfo {author} {\bibfnamefont {U.-K.}\ \bibnamefont {Yang}},\ }\href
  {\doibase 10.1016/j.nuclphysbps.2004.11.208} {\bibfield  {journal} {\bibinfo
  {journal} {Nucl.~Phys.~Proc.~Suppl.}\ }\textbf {\bibinfo {volume} {139}},\
  \bibinfo {pages} {113} (\bibinfo {year} {2005})},\ \Eprint
  {http://arxiv.org/abs/hep-ph/0411202} {arXiv:hep-ph/0411202} \BibitemShut
  {NoStop}%
\bibitem [{\citenamefont {Rodrigues}\ \emph
  {et~al.}(2016{\natexlab{b}})\citenamefont {Rodrigues}, \citenamefont
  {Wilkinson},\ and\ \citenamefont {McFarland}}]{Rodrigues:2016xjj}%
  \BibitemOpen
  \bibfield  {author} {\bibinfo {author} {\bibfnamefont {P.}~\bibnamefont
  {Rodrigues}}, \bibinfo {author} {\bibfnamefont {C.}~\bibnamefont
  {Wilkinson}}, \ and\ \bibinfo {author} {\bibfnamefont {K.}~\bibnamefont
  {McFarland}},\ }\href {\doibase 10.1140/epjc/s10052-016-4314-3} {\bibfield
  {journal} {\bibinfo  {journal} {Eur. Phys. J.}\ }\textbf {\bibinfo {volume}
  {C76}},\ \bibinfo {pages} {474} (\bibinfo {year} {2016}{\natexlab{b}})},\
  \Eprint {http://arxiv.org/abs/1601.01888} {arXiv:1601.01888} \BibitemShut
  {NoStop}%
\bibitem [{\citenamefont {Wilkinson}\ \emph {et~al.}(2014)\citenamefont
  {Wilkinson} \emph {et~al.}}]{Wilkinson:2014yfa}%
  \BibitemOpen
  \bibfield  {author} {\bibinfo {author} {\bibfnamefont {C.}~\bibnamefont
  {Wilkinson}} \emph {et~al.},\ }\href {\doibase 10.1103/PhysRevD.90.112017}
  {\bibfield  {journal} {\bibinfo  {journal} {Phys.~Rev.~D}\ }\textbf {\bibinfo
  {volume} {90}},\ \bibinfo {pages} {112017} (\bibinfo {year} {2014})},\
  \Eprint {http://arxiv.org/abs/1411.4482} {arXiv:1411.4482} \BibitemShut
  {NoStop}%
\bibitem [{\citenamefont {Higuera}\ \emph {et~al.}(2014)\citenamefont {Higuera}
  \emph {et~al.}}]{Higuera:2014azj}%
  \BibitemOpen
  \bibfield  {author} {\bibinfo {author} {\bibfnamefont {A.}~\bibnamefont
  {Higuera}} \emph {et~al.} (\bibinfo {collaboration} {MINERvA
  Collaboration}),\ }\href {\doibase 10.1103/PhysRevLett.113.261802} {\bibfield
   {journal} {\bibinfo  {journal} {Phys. Rev. Lett.}\ }\textbf {\bibinfo
  {volume} {113}},\ \bibinfo {pages} {261802} (\bibinfo {year} {2014})},\
  \Eprint {http://arxiv.org/abs/1409.3835} {arXiv:1409.3835} \BibitemShut
  {NoStop}%
\bibitem [{\citenamefont {Mislivec}\ \emph {et~al.}(2018)\citenamefont
  {Mislivec} \emph {et~al.}}]{Mislivec:2017qfz}%
  \BibitemOpen
  \bibfield  {author} {\bibinfo {author} {\bibfnamefont {A.}~\bibnamefont
  {Mislivec}} \emph {et~al.} (\bibinfo {collaboration} {MINERvA
  Collaboration}),\ }\href {\doibase 10.1103/PhysRevD.97.032014} {\bibfield
  {journal} {\bibinfo  {journal} {Phys.~Rev.~D}\ }\textbf {\bibinfo {volume}
  {97}},\ \bibinfo {pages} {032014} (\bibinfo {year} {2018})},\ \Eprint
  {http://arxiv.org/abs/1711.01178} {arXiv:1711.01178} \BibitemShut {NoStop}%
\bibitem [{\citenamefont {Berger}\ and\ \citenamefont
  {Sehgal}(2009)}]{Berger:2008xs}%
  \BibitemOpen
  \bibfield  {author} {\bibinfo {author} {\bibfnamefont {C.}~\bibnamefont
  {Berger}}\ and\ \bibinfo {author} {\bibfnamefont {L.}~\bibnamefont
  {Sehgal}},\ }\href {\doibase 10.1103/PhysRevD.79.053003} {\bibfield
  {journal} {\bibinfo  {journal} {Phys.~Rev.~D}\ }\textbf {\bibinfo {volume}
  {79}},\ \bibinfo {pages} {053003} (\bibinfo {year} {2009})},\ \Eprint
  {http://arxiv.org/abs/0812.2653} {arXiv:0812.2653} \BibitemShut {NoStop}%
\bibitem [{\citenamefont {Gran}\ \emph {et~al.}(2013)\citenamefont {Gran},
  \citenamefont {Nieves}, \citenamefont {Sanchez},\ and\ \citenamefont
  {Vicente~Vacas}}]{Gran:2013kda}%
  \BibitemOpen
  \bibfield  {author} {\bibinfo {author} {\bibfnamefont {R.}~\bibnamefont
  {Gran}}, \bibinfo {author} {\bibfnamefont {J.}~\bibnamefont {Nieves}},
  \bibinfo {author} {\bibfnamefont {F.}~\bibnamefont {Sanchez}}, \ and\
  \bibinfo {author} {\bibfnamefont {M.}~\bibnamefont {Vicente~Vacas}},\ }\href
  {\doibase 10.1103/PhysRevD.88.113007} {\bibfield  {journal} {\bibinfo
  {journal} {Phys.~Rev.~D}\ }\textbf {\bibinfo {volume} {88}},\ \bibinfo
  {pages} {113007} (\bibinfo {year} {2013})},\ \Eprint
  {http://arxiv.org/abs/1307.8105} {arXiv:1307.8105} \BibitemShut {NoStop}%
\bibitem [{\citenamefont {Gran}(2017)}]{Gran:2017psn}%
  \BibitemOpen
  \bibfield  {author} {\bibinfo {author} {\bibfnamefont {R.}~\bibnamefont
  {Gran}},\ }\href@noop {} {\  (\bibinfo {year} {2017})},\ \Eprint
  {http://arxiv.org/abs/1705.02932} {arXiv:1705.02932} \BibitemShut {NoStop}%
\bibitem [{\citenamefont {Rodrigues}\ \emph {et~al.}()\citenamefont {Rodrigues}
  \emph {et~al.}}]{minerva2p2h:2019}%
  \BibitemOpen
  \bibfield  {author} {\bibinfo {author} {\bibfnamefont {P.~A.}\ \bibnamefont
  {Rodrigues}} \emph {et~al.} (\bibinfo {collaboration} {MINERvA
  Collaboration}),\ }\href@noop {} {\bibinfo  {journal} {in preparation}\
  }\BibitemShut {NoStop}%
\bibitem [{\citenamefont {Betancourt}\ \emph {et~al.}(2017)\citenamefont
  {Betancourt} \emph {et~al.}}]{Betancourt:2017uso}%
  \BibitemOpen
\bibfield  {journal} {  }\bibfield  {author} {\bibinfo {author} {\bibfnamefont
  {M.}~\bibnamefont {Betancourt}} \emph {et~al.} (\bibinfo {collaboration}
  {MINERvA Collaboration}),\ }\href {\doibase 10.1103/PhysRevLett.119.082001}
  {\bibfield  {journal} {\bibinfo  {journal} {Phys. Rev. Lett.}\ }\textbf
  {\bibinfo {volume} {119}},\ \bibinfo {pages} {082001} (\bibinfo {year}
  {2017})},\ \Eprint {http://arxiv.org/abs/1705.03791} {arXiv:1705.03791}
  \BibitemShut {NoStop}%
\bibitem [{\citenamefont {Altinok}\ \emph {et~al.}(2017)\citenamefont {Altinok}
  \emph {et~al.}}]{Altinok:2017xua}%
  \BibitemOpen
  \bibfield  {author} {\bibinfo {author} {\bibfnamefont {O.}~\bibnamefont
  {Altinok}} \emph {et~al.} (\bibinfo {collaboration} {MINERvA
  Collaboration}),\ }\href {\doibase 10.1103/PhysRevD.96.072003} {\bibfield
  {journal} {\bibinfo  {journal} {Phys.~Rev.~D}\ }\textbf {\bibinfo {volume}
  {96}},\ \bibinfo {pages} {072003} (\bibinfo {year} {2017})},\ \Eprint
  {http://arxiv.org/abs/1708.03723} {arXiv:1708.03723} \BibitemShut {NoStop}%
\bibitem [{\citenamefont {Patrick}\ \emph {et~al.}(2018)\citenamefont {Patrick}
  \emph {et~al.}}]{Patrick:2018gvi}%
  \BibitemOpen
  \bibfield  {author} {\bibinfo {author} {\bibfnamefont {C.~E.}\ \bibnamefont
  {Patrick}} \emph {et~al.} (\bibinfo {collaboration} {MINERvA
  Collaboration}),\ }\href {\doibase 10.1103/PhysRevD.97.052002} {\bibfield
  {journal} {\bibinfo  {journal} {Phys.~Rev.~D}\ }\textbf {\bibinfo {volume}
  {97}},\ \bibinfo {pages} {052002} (\bibinfo {year} {2018})},\ \Eprint
  {http://arxiv.org/abs/1801.01197} {arXiv:1801.01197} \BibitemShut {NoStop}%
\bibitem [{\citenamefont {Meyer}\ \emph {et~al.}(2016)\citenamefont {Meyer},
  \citenamefont {Betancourt}, \citenamefont {Gran},\ and\ \citenamefont
  {Hill}}]{Meyer:2016oeg}%
  \BibitemOpen
  \bibfield  {author} {\bibinfo {author} {\bibfnamefont {A.~S.}\ \bibnamefont
  {Meyer}}, \bibinfo {author} {\bibfnamefont {M.}~\bibnamefont {Betancourt}},
  \bibinfo {author} {\bibfnamefont {R.}~\bibnamefont {Gran}}, \ and\ \bibinfo
  {author} {\bibfnamefont {R.~J.}\ \bibnamefont {Hill}},\ }\href {\doibase
  10.1103/PhysRevD.93.113015} {\bibfield  {journal} {\bibinfo  {journal}
  {Phys.~Rev.~D}\ }\textbf {\bibinfo {volume} {93}},\ \bibinfo {pages} {113015}
  (\bibinfo {year} {2016})},\ \Eprint {http://arxiv.org/abs/1603.03048}
  {arXiv:1603.03048 [hep-ph]} \BibitemShut {NoStop}%
\bibitem [{\citenamefont {Valverde}\ \emph {et~al.}(2006)\citenamefont
  {Valverde}, \citenamefont {Amaro},\ and\ \citenamefont
  {Nieves}}]{Valverde:2006zn}%
  \BibitemOpen
  \bibfield  {author} {\bibinfo {author} {\bibfnamefont {M.}~\bibnamefont
  {Valverde}}, \bibinfo {author} {\bibfnamefont {J.~E.}\ \bibnamefont {Amaro}},
  \ and\ \bibinfo {author} {\bibfnamefont {J.}~\bibnamefont {Nieves}},\ }\href
  {\doibase 10.1016/j.physletb.2006.05.053} {\bibfield  {journal} {\bibinfo
  {journal} {Phys. Lett.}\ }\textbf {\bibinfo {volume} {B638}},\ \bibinfo
  {pages} {325} (\bibinfo {year} {2006})},\ \Eprint
  {http://arxiv.org/abs/hep-ph/0604042} {arXiv:hep-ph/0604042} \BibitemShut
  {NoStop}%
\bibitem [{\citenamefont {Martini}\ \emph {et~al.}(2016)\citenamefont
  {Martini}, \citenamefont {Jachowicz}, \citenamefont {Ericson}, \citenamefont
  {Pandey}, \citenamefont {Van~Cuyck},\ and\ \citenamefont
  {Van~Dessel}}]{Martini:2016eec}%
  \BibitemOpen
  \bibfield  {author} {\bibinfo {author} {\bibfnamefont {M.}~\bibnamefont
  {Martini}}, \bibinfo {author} {\bibfnamefont {N.}~\bibnamefont {Jachowicz}},
  \bibinfo {author} {\bibfnamefont {M.}~\bibnamefont {Ericson}}, \bibinfo
  {author} {\bibfnamefont {V.}~\bibnamefont {Pandey}}, \bibinfo {author}
  {\bibfnamefont {T.}~\bibnamefont {Van~Cuyck}}, \ and\ \bibinfo {author}
  {\bibfnamefont {N.}~\bibnamefont {Van~Dessel}},\ }\href {\doibase
  10.1103/PhysRevC.94.015501} {\bibfield  {journal} {\bibinfo  {journal}
  {Phys.~Rev.~C}\ }\textbf {\bibinfo {volume} {94}},\ \bibinfo {pages} {015501}
  (\bibinfo {year} {2016})},\ \Eprint {http://arxiv.org/abs/1602.00230}
  {arXiv:1602.00230} \BibitemShut {NoStop}%
\bibitem [{\citenamefont {Aguilar-Arevalo}\ \emph
  {et~al.}(2011{\natexlab{a}})\citenamefont {Aguilar-Arevalo} \emph
  {et~al.}}]{AguilarArevalo:2010bm}%
  \BibitemOpen
  \bibfield  {author} {\bibinfo {author} {\bibfnamefont {A.~A.}\ \bibnamefont
  {Aguilar-Arevalo}} \emph {et~al.} (\bibinfo {collaboration} {MiniBooNE
  Collaboration}),\ }\href {\doibase 10.1103/PhysRevD.83.052007} {\bibfield
  {journal} {\bibinfo  {journal} {Phys.~Rev.~D}\ }\textbf {\bibinfo {volume}
  {83}},\ \bibinfo {pages} {052007} (\bibinfo {year} {2011}{\natexlab{a}})},\
  \Eprint {http://arxiv.org/abs/1011.3572} {arXiv:1011.3572} \BibitemShut
  {NoStop}%
\bibitem [{\citenamefont {Aguilar-Arevalo}\ \emph
  {et~al.}(2011{\natexlab{b}})\citenamefont {Aguilar-Arevalo} \emph
  {et~al.}}]{AguilarArevalo:2010xt}%
  \BibitemOpen
  \bibfield  {author} {\bibinfo {author} {\bibfnamefont {A.~A.}\ \bibnamefont
  {Aguilar-Arevalo}} \emph {et~al.} (\bibinfo {collaboration} {MiniBooNE
  Collaboration}),\ }\href {\doibase 10.1103/PhysRevD.83.052009} {\bibfield
  {journal} {\bibinfo  {journal} {Phys.~Rev.~D}\ }\textbf {\bibinfo {volume}
  {83}},\ \bibinfo {pages} {052009} (\bibinfo {year} {2011}{\natexlab{b}})},\
  \Eprint {http://arxiv.org/abs/1010.3264} {arXiv:1010.3264} \BibitemShut
  {NoStop}%
\bibitem [{\citenamefont {McGivern}\ \emph {et~al.}(2016)\citenamefont
  {McGivern} \emph {et~al.}}]{McGivern:2016bwh}%
  \BibitemOpen
  \bibfield  {author} {\bibinfo {author} {\bibfnamefont {C.~L.}\ \bibnamefont
  {McGivern}} \emph {et~al.} (\bibinfo {collaboration} {MINERvA
  Collaboration}),\ }\href {\doibase 10.1103/PhysRevD.94.052005} {\bibfield
  {journal} {\bibinfo  {journal} {Phys.~Rev.~D}\ }\textbf {\bibinfo {volume}
  {94}},\ \bibinfo {pages} {052005} (\bibinfo {year} {2016})},\ \Eprint
  {http://arxiv.org/abs/1606.07127} {arXiv:1606.07127} \BibitemShut {NoStop}%
\bibitem [{\citenamefont {Adamson}\ \emph {et~al.}(2015)\citenamefont {Adamson}
  \emph {et~al.}}]{Adamson:2014pgc}%
  \BibitemOpen
  \bibfield  {author} {\bibinfo {author} {\bibfnamefont {P.}~\bibnamefont
  {Adamson}} \emph {et~al.} (\bibinfo {collaboration} {MINOS Collaboration}),\
  }\href {\doibase 10.1103/PhysRevD.91.012005} {\bibfield  {journal} {\bibinfo
  {journal} {Phys.~Rev.~D}\ }\textbf {\bibinfo {volume} {91}},\ \bibinfo
  {pages} {012005} (\bibinfo {year} {2015})},\ \Eprint
  {http://arxiv.org/abs/1410.8613} {arXiv:1410.8613} \BibitemShut {NoStop}%
\bibitem [{\citenamefont {D'Agostini}(1995)}]{D'Agostini:1994zf}%
  \BibitemOpen
  \bibfield  {author} {\bibinfo {author} {\bibfnamefont {G.}~\bibnamefont
  {D'Agostini}},\ }\href {\doibase 10.1016/0168-9002(95)00274-X} {\bibfield
  {journal} {\bibinfo  {journal} {Nucl.~Instrum.~Meth.~A}\ }\textbf {\bibinfo
  {volume} {362}},\ \bibinfo {pages} {487} (\bibinfo {year}
  {1995})}\BibitemShut {NoStop}%
\bibitem [{\citenamefont {D'Agostini}(2010)}]{DAgostini:2010xxxxx}%
  \BibitemOpen
  \bibfield  {author} {\bibinfo {author} {\bibfnamefont {G.}~\bibnamefont
  {D'Agostini}},\ }\href@noop {} {\  (\bibinfo {year} {2010})},\ \Eprint
  {http://arxiv.org/abs/1010.0632} {arXiv:1010.0632} \BibitemShut {NoStop}%
\bibitem [{\citenamefont {Adye}()}]{Adye:2011gm}%
  \BibitemOpen
  \bibfield  {author} {\bibinfo {author} {\bibfnamefont {T.}~\bibnamefont
  {Adye}},\ }\href@noop {} {\ }\Eprint {http://arxiv.org/abs/1105.1160}
  {arXiv:1105.1160} \BibitemShut {NoStop}%
\end{thebibliography}%



\end{document}